\documentclass[opre,nonblindrev]{informs3_hide}

\DoubleSpacedXI 

\usepackage{endnotes,bbm,multirow,multicol,float,wrapfig,algorithm,enumitem,algpseudocode,xspace}

\newenvironment{pseudocode}[1][htb]{
	\floatname{algorithm}{Algorithm}
	\begin{algorithm}[#1]%
	}{\end{algorithm}}

\newcommand{\bE}{\mathbb{E}}
\newcommand{\bN}{\mathbb{N}}

\newcommand{\bR}{\mathbb{R}}

\newcommand{\cA}{\mathcal{A}}
\newcommand{\cF}{\mathcal{F}}
\newcommand{\cP}{\mathcal{P}}
\newcommand{\cS}{\mathcal{S}}

\newcommand{\ximax}{\xi_{\max}}
\newcommand{\kmin}{k_{\min}}
\newcommand{\ALG}{\mathsf{ALG}}
\newcommand{\OPT}{\mathsf{OPT}}
\newcommand{\rL}{r^{(\mathsf{L})}}
\newcommand{\rH}{r^{(\mathsf{H})}}
\newcommand{\balance}{\textsc{Multi-price Balance}\xspace}
\newcommand{\ranking}{\textsc{Multi-price Ranking}\xspace}

\newcommand{\tCR}{\tilde{F}}
\newcommand{\tL}{\tilde{L}}
\newcommand{\tPhi}{\tilde{\Phi}}
\newcommand{\tell}{\tilde{\ell}}

\newcommand{\available}{\mathsf{available}}
\newcommand{\wcrit}{W^{\mathsf{crit}}}
\newcommand{\zcrit}{Z^{\mathsf{crit}}}

\newcommand{\Rmin}{r^{\min}}
\newcommand{\Rmax}{r^{\max}}

\newcommand{\wdt}{5}
\newcommand{\hgt}{5}
\newcommand{\wdtt}{5}
\newcommand{\legend}{3}
\newcommand{\hgtt}{3}
\newcommand{\xAxisColor}{Blue}
\newcommand{\yAxisColor}{Red}
\newcommand{\zAxisColor}{Blue}
\newcommand{\axisLabelDist}{0.4}

\newcommand{\legendDist}{0.5}

\newcommand{\horSc}{4}

\usepackage{natbib}
 \bibpunct[, ]{(}{)}{,}{a}{}{,}%
 \def\bibfont{\small}%
\restylefloat{figure}
\usepackage[dvipsnames]{xcolor}
\usepackage{tikz}
\usetikzlibrary{arrows}

\TheoremsNumberedThrough     
\ECRepeatTheorems

\EquationsNumberedThrough    

\MANUSCRIPTNO{OPRE-2017-09-472.R1} 

\begin{document}


 \RUNAUTHOR{Ma and Simchi-Levi}

\RUNTITLE{Tight Weight-dependent Competitive Ratios for Online Matching, Assortment, and Pricing}

\TITLE{Algorithms for Online Matching, Assortment, and Pricing with Tight Weight-dependent Competitive Ratios}

\ARTICLEAUTHORS{%
\AUTHOR{Will Ma}
\AFF{Operations Research Center, Massachusetts Institute of Technology, Cambridge, MA 02139, \EMAIL{willma@mit.edu}} 
\AUTHOR{David Simchi-Levi}
\AFF{Institute for Data, Systems, and Society, Department of Civil and Environmental Engineering, and Operations Research Center, Massachusetts Institute of Technology, Cambridge, MA 02139, \EMAIL{dslevi@mit.edu}}
}
\ABSTRACT{

Motivated by the dynamic assortment offerings and item pricings occurring in e-commerce, we study a general problem of allocating finite inventories to heterogeneous customers arriving sequentially.  We analyze this problem under the framework of competitive analysis, where the sequence of customers is unknown and does not necessarily follow any pattern.  Previous work in this area, studying online matching, advertising, and assortment problems, has focused on the case where each item can only be sold at a single price, resulting in algorithms which achieve the best-possible competitive ratio of 1-1/e.

In this paper, we extend all of these results to allow for items having multiple feasible prices.  Our algorithms achieve the best-possible weight-dependent competitive ratios, which depend on the sets of feasible prices given in advance.  Our algorithms are also simple and intuitive; they are based on constructing a class of universal ``value functions'' which integrate the selection of items and prices offered.

Finally, we test our algorithms on the publicly-available hotel data set of Bodea et al. (2009), where there are multiple items (hotel rooms) each with multiple prices (fares at which the room could be sold).  We find that applying our algorithms, as a ``hybrid'' with algorithms which attempt to forecast and learn the future transactions, results in the best performance.

}



\maketitle

\section{Introduction}

In this paper we study a general online resource allocation problem, motivated by dynamic assortment and pricing in revenue management.
Consider an airline website selling parallel flights, i.e.\ different flights which depart
from the same origin to the same destination around the same time.
Each flight corresponds to an \textit{item} which could be sold, and its seat capacity corresponds to the unreplenishable \textit{starting inventory} of that item.
Each flight has multiple fare classes (e.g.\ Economy, Basic Economy) which correspond to \textit{prices} at which that item could be sold.
We will refer to this collection of initial information (items, inventories, prices) as the \textit{setup}.

Over the booking horizon, heterogeneous customers sequentially arrive to the airline's website.
We assume that the airline can reliably estimate each customer's \textit{choice probabilities} from historical data.
That is, upon a customer's arrival, for any combination of items and prices that could be shown, the stochastic distribution of how the customer would choose among those items/prices is given.
The customer is assumed to choose at most one item and one price, as the flights are parallel.
The customer could also choose to make no purchase.
Given the choice probabilities, the airline selects an \textit{assortment} of items and corresponding prices to show the customer, where items with zero remaining inventory cannot be shown.
The customer's decision is realized immediately afterward, and if she makes a purchase, then the airline earns the corresponding price as revenue, and depletes one unit of inventory of the corresponding item.
The airline wants to maximize its cumulative revenue earned before the booking horizon is over (or all the flights are full).

We study this problem under the framework of competitive analysis, where the sequence of customers to arrive over the booking horizon is unknown and does not necessarily follow any pattern.
Instead, the airline seeks to have a good relative performance on all possible sequences.
It offers assortments and pricings using a (possibly randomized) \textit{online algorithm}, which can make decisions based on only the setup and the arrival sequence/purchase realizations seen so far.
For $c\le1$, the online algorithm is said to be $c$-\textit{competitive}, or \textit{achieve a competitive ratio of} $c$, if
\begin{align} \label{eqn::introCR}
\inf_{\text{arrival sequences }\cA}\frac{\bE[\ALG(\cA)]}{\OPT(\cA)}\ge c,
\end{align}
where $\bE[\ALG(\cA)]$ denotes the algorithm's expected revenue on arrival sequence $\cA$, and $\OPT(\cA)$ denotes the value of an optimum which knows the entirety of $\cA$ in advance.
In this paper, we will allow the competitive ratio guarantee to be \textit{setup-dependent}; that is, the value of $c$ in (\ref{eqn::introCR}) can be a function of the items, their starting inventories, and prices.
We are interested in online algorithms which achieve the best-possible competitive ratios for various families of setups.

\subsection{Overview of Result, and Relation to Previous Results}

For setups where each item has a single fare class, the competitive ratio of the above problem has been analyzed extensively under many streams of literature, which we review below.
\begin{enumerate}
\item \textbf{Online Assortment}:
If each item has a single price, then the above problem formulation is exactly the online assortment problem of \citet{GNR14}.
The authors use an algorithm which judiciously ``balances'' between offering different items, based on their remaining inventory levels.  They show, among other results, that their algorithm is $(1-1/e)$-competitive in the \textit{asymptotic regime}.
That is, their value of $c$ in (\ref{eqn::introCR}) depends on the smallest starting inventory amount in the setup, and approaches $1-1/e$ as all starting inventories approach $\infty$.
Without large starting inventories, this problem has also been studied in the special case where all offered assortments must have size 1, in which case it becomes the \textit{online matching with stochastic rewards} problem \citep{MP12,MWZ14}.

\item \textbf{Online Vertex-weighted Matching}: 
Consider the special case of the problem where the outcome of any assortment offering is deterministic, and given upon the customer's arrival.
In this case, we know the maximum (possibly 0) a customer is willing to pay for each item, and our decision can be reduced to selecting an item to offer to the customer at her maximum-willingness-to-pay (we can also offer no item).
We will refer to this problem as the \textit{deterministic case}; it can be viewed as an online weighted matching problem.

If each item is restricted to have a single price, then we get the \textit{online vertex-weighted matching} problem of \citet{AGKM11}.
The authors develop an algorithm which randomly ``ranks'' the items and matches higher-ranked items first, and show that it is $(1-1/e)$-competitive.
Their result generalizes the classical result of \citet{KVV90} for the unweighted online bipartite matching problem (where all items have the \textit{same} price).
\item \textbf{Adwords}: The Adwords problem of \citet{MSVV07} is central to online advertising and features budget-constrained bidders instead of inventory-constrained items.  It also uses the idea of ``balancing'', between the bidders' budgets in this case, to achieve $(1-1/e)$-competitiveness in an asymptotic \textit{small bids} regime.
Although its budget constraints are not directly captured by our model, we show that in this asymptotic regime, the Adwords problem corresponds to a version of our problem where each item has a \textit{single} price (despite each bidder having multiple bid values---we explain the reduction in Section~\ref{sect::generalizations}).
\end{enumerate}

Our main contribution, motivated by the parallel flights problem, is extending all of the preceding results to setups where items could have multiple prices, that are known in advance.
Note that such a setup also arises naturally from models where the customers have been classified into ``types'', and there is a ``match quality'' score between each item and each type---in that case, the price set of an item consists of the item's possible match scores.

However, allowing for multiple prices per item runs into a known impossibility result: even in the deterministic case, which corresponds to the aforementioned online weighted matching problem, it is not possible to provide a non-zero competitive ratio guarantee $c$ which holds for every multi-price setup (the way $c=1-1/e$ was a constant guarantee for every single-price setup).
This is because the moment we commit to a match, unboundedly larger edge weights can arrive afterward---see \citet[Ch.~7]{Meh13}.
Therefore, previous work in online weighted matching has assumed that matches can be freely disposed if larger weights arrive later \citep{FKMMP09}, or that arrivals appear in a random order \citep{KRTV13}.

In our paper, we instead assume that the price sets (i.e., the possible edge weights) are known in advance, and derive \textit{weight-dependent} competitive ratio guarantees, where our value of $c$ in (\ref{eqn::introCR}) will depend on the setup; specifically, the price sets $\cP_1,\ldots,\cP_n$ of the $n$ items.
Our algorithms also make use of the knowledge of the price sets.
Our competitive ratio results are based on establishing a universal mapping $F$ from price sets $\cP$ to ratios in $[0,1-1/e]$, such that:
\begin{enumerate}
\item Our \balance algorithm, which extends the existing ``inventory balancing'' algorithms, is 
$\min_iF(\cP_i)$-competitive in the asymptotic regime;
\item Our \ranking algorithm, which extends the existing ``randomized ranking'' algorithm, is $\min_iF(\cP_i)$-competitive in the deterministic case;
\item Any (deterministic or randomized) algorithm can be at most $\min_iF(\cP_i)$-competitive for the family of setups with price sets chosen from $\cP_1,\ldots,\cP_n$, even if we restrict the setups to have asymptotic starting inventories and/or deterministic arrival sequences.
\end{enumerate}
For any singleton price set $\cP$,
$F(\cP)=1-1/e$,
and hence if $|\cP_1|=\ldots=|\cP_n|=1$, then our results recover existing results: the $(1-1/e)$-competitiveness of inventory balancing in the asymptotic regime, the $(1-1/e)$-competitiveness of randomized ranking in the deterministic case, and a single counterexample which shows that both of these algorithms are tight.
$F(\cP)$ approaches 0 if $\cP$ contains both a large number of prices and large ratios between its prices, so our statement~3 also recovers the known impossibility result.


\subsection{A Bid Price Algorithm when Items have Multiple Prices}\label{subsect::introBidPrice}

We illustrate the necessity for our new algorithms by comparing \balance to the existing inventory balancing algorithm of \citet{GNR14};
similar arguments can be made in comparisons to the existing online vertex-weighted matching and Adwords algorithms.

Suppose there are parallel flights, whose seats have the same two fare classes: a lower price of $\rL=150$, and a higher price of $\rH=450$.
At any point in time, for each flight $i$, let $w_i$ denote the fraction its starting inventory which has been sold.
The algorithm of \citet{GNR14} would associate each fare class $j\in\{\mathsf{L},\mathsf{H}\}$ of each flight $i$ with a ``pseudorevenue'' equal to
\begin{align} \label{eqn::pseudorevenueMult}
r^{(j)}\cdot\Psi(w_i),
\end{align}
where $\Psi$ is a decreasing function that \textit{penalizes} the revenues associated with flights $i$ which are almost full.
The algorithm then offers, to each customer, the assortment which maximizes the expected pseudorevenue of the (flight, fare)-combination that the customer would choose.

In (\ref{eqn::pseudorevenueMult}), although $\Psi(w_i)$ will disincentivize the offering of a flight $i$ whose $w_i$ is large, the algorithm has no way of setting a ``booking limit''---preventing sales at the lower price $\rL$ while still allowing sales at
price $\rH$.
Given a stream of customers who are only interested in the lower price, the algorithm would sell all the seats at price $\rL$, without realizing the opportunity cost of $\rH$ it gave up.
Since this could happen to every flight $i$, the algorithm's competitiveness is at most $\frac{\rL}{\rH}$.

To improve upon this, our algorithm must implement some notion of ``booking limits''.
As a result, we define the pseudorevenue associated with fare class $j$ of flight $i$ to be
\begin{align} \label{eqn::pseudorevenueAdditive}
r^{(j)}-\Phi(w_i),
\end{align}
where $\Phi$ is an increasing function that sets a \textit{cost} to selling flights $i$ which are almost full.
\balance uses the same idea of maximizing expected pseudorevenue, with the modification that it rejects the customer outright if the
maximum
expected pseudorevenue is non-positive.

\begin{figure}
\centering\begin{tikzpicture}
\draw(0,0)node[below left]{0}--(\horSc,0)node[below]{1};
\node at (\horSc/2,-1){Fraction Sold, $w_i$};
\node at (-3/2,3/2){$\Phi(w_i)$};
\draw(0,0)--(0,3)node[left]{\$450};
\draw(0.63*\horSc,1)--(0.63*\horSc,0)node[below]{$\alpha$}[dashed];
\draw(0.63*\horSc,1)--(0,1)node[left]{\$150}[dashed];
\draw(\horSc,3)--(\horSc,0)[dashed];
\draw(\horSc,3)--(0,3)[dashed];
\draw[color=red,domain=0:0.63*\horSc,samples=100,ultra thick]plot(\x,{(exp(\x/\horSc)-1)/(exp(0.63)-1)});
\draw[color=red,domain=0.63*\horSc:\horSc,samples=100,ultra thick]plot(\x,{1+2*(exp(\x/\horSc-0.63)-1)/(exp(0.37)-1)});
\end{tikzpicture}
\caption{The $\Phi$ function for an item with feasible price set $\{150,450\}$.}
\label{fig::valueFn}
\end{figure}
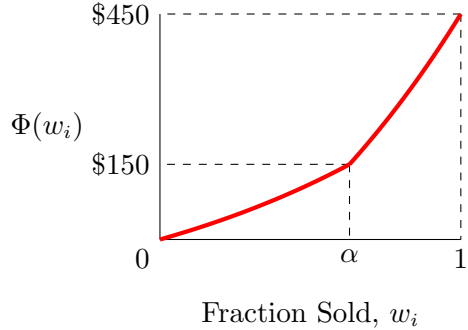

The exact form of function $\Phi$ for this example is shown in Figure~\ref{fig::valueFn}.
Note that for a flight $i$, if the fraction sold $w_i$ is between $\alpha$ and 1, then the pseudorevenue~(\ref{eqn::pseudorevenueAdditive}) will be negative for the lower price of 150 and positive for the higher price 450, producing a desirable ``booking limit'' at $w_i=\alpha$.
Other than that, $\Phi$ still produces a continuously-increasing cost as $w_i$ increases from 0 to 1, allowing us to trade off between offering the different flights $i$ based on their values of $w_i$.

In (\ref{eqn::pseudorevenueAdditive}), $\Phi(w_i)$ can be interpreted as a \textit{bid price}, or the value placed on one unit of item $i$'s inventory.
Optimizing based on bid prices is a classical idea in revenue management (see \citet{TvR06,LvR08}), where typically the bid prices are computed using a large LP which encompasses both the inventories and the forecasted distribution of future customers.
However, since we make no assumptions about future customers, our bid prices are based on only the remaining inventories, like the balance algorithms from competitive analysis.

\subsection{Our Competitive Ratio Guarantees}

In general, for any price set $\cP$ we define a construction $\Phi_{\cP}$ which we call a \textit{value function}.
In a setup, if $\cP_1,\ldots,\cP_n$ denotes the price sets of the items, then
\balance defines the pseudorevenues of each item $i$ using value function $\Phi_{\cP_i}$ in expression~(\ref{eqn::pseudorevenueAdditive}).
Our \ranking algorithm
uses
the same value functions, but applies them to a ``random seed'' instead.

Note that for each item $i$, the construction of $\Phi_{\cP_i}$ from $\cP_i$ is universal in that it does not depend on other parameters in the setup, e.g.\ the price sets of the other items.
Our mapping $F$ from price sets to ratios was also universal.
The fact that separately determining the value function $\Phi_{\cP_i}$ for each item $i$ leads to the best-possible competitive ratio of $\min_iF(\cP_i)$ is, in our opinion, very surprising---see also the discussions in \citet{DJ12,DJK13}.
For any $\cP$, our exact derivation of $\Phi_{\cP}$ and $F(\cP)$ comes from the solution of a differential equation, which arises from a primal-dual analysis based on \citet{BJN07}.

When $|\cP|=1$, with $\cP=\{r\}$, value function $\Phi_{\cP}(w)=r(1-\Psi(w))$, and hence our notion of pseudorevenue in (\ref{eqn::pseudorevenueAdditive}) coincides with the existing notion in (\ref{eqn::pseudorevenueMult}).  If each item has a single price, then our algorithms will coincide with the existing ones.

We now give a flavor of our new results with $|\cP|=2$.
Let $\cP=\{r,\xi r\}$, where $\xi>1$ is the ratio from high price to low price.
The value function $\Phi_{\cP}$ depends on $\xi$ (see Figure~\ref{fig::valueFn} for an example with $\xi=\frac{450}{150}=3$).  For any $\xi$, $F(\cP)$ equals
\begin{equation}\label{eqn::intro}
1-\frac{\sqrt{1+4\xi(\xi-1)/e}-1}{2(\xi-1)}=:F(\xi),
\end{equation}
and the booking limit implied by $\Phi_{\cP}$ (i.e.\ the corresponding value of $\alpha$ in Figure~\ref{fig::valueFn}) equals $\ln(\frac{1}{1-F(\xi)})$.
We note that this is different from the booking limit of $\frac{\xi}{2\xi-1}$ derived by \citet{BQ09}, which is optimal for selling a \textit{single} item whose price set is 
$\{r,\xi r\}$.
For any value of $\xi$, the booking limit implied by our function $\Phi_{\{r,\xi r\}}$ is greater than $\frac{\xi}{2\xi-1}$, which means that our algorithm is willing to sell a greater fraction of units at the lower price.
The intuitive explanation of this is that with multiple items, there is less upside to reserving inventory for higher prices, because the reserved units still have to compete with other items to be sold.

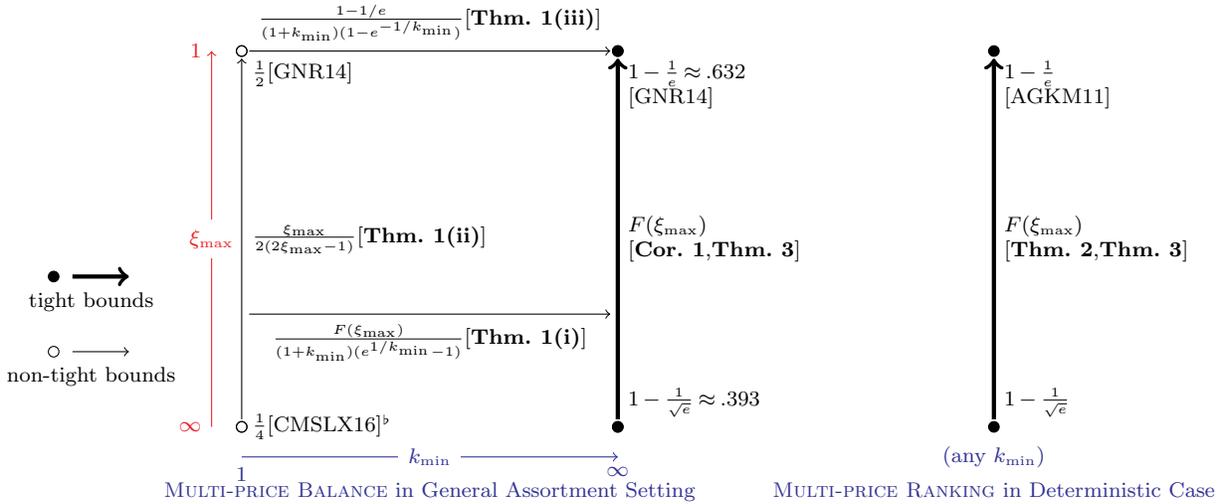
\begin{figure}
\begin{center}
\begin{tikzpicture}[font=\scriptsize]
\draw(-\axisLabelDist,0)node[left,\yAxisColor]{$\infty$};
\draw[->](-\axisLabelDist,0)--(-\axisLabelDist,\hgt)[\yAxisColor]node[midway,fill=white]{$\ximax$};
\draw(-\axisLabelDist,\hgt)node[left,\yAxisColor]{1};

\draw(0,-\axisLabelDist)node[below,\xAxisColor]{1};
\draw[->](0,-\axisLabelDist)--(\wdt,-\axisLabelDist)[\xAxisColor]node[midway,fill=white]{$\kmin$};
\draw(\wdt,-\axisLabelDist)node[below,\xAxisColor]{$\infty$};
\draw(\wdt/2,-1.5*\axisLabelDist)node[below,\xAxisColor]{\balance in General Assortment Setting};

\draw(0,0)circle(2pt);
\draw(\wdt,0)circle(2pt);
\draw(0,\hgt)circle(2pt);
\draw(\wdt,\hgt)circle(2pt);

\filldraw(\legendDist-\legend,\hgtt-1)circle(2pt);
\draw[->](\legendDist+0.25-\legend,\hgtt-1)--(\legendDist+1-\legend,\hgtt-1)[ultra thick];
\node at (\legendDist+0.5-\legend,\hgtt-1-0.33){tight bounds};
\draw(\legendDist-\legend,\hgtt-2)circle(2pt);
\draw[->](\legendDist+0.25-\legend,\hgtt-2)--(\legendDist+1-\legend,\hgtt-2);
\node at (\legendDist+0.5-\legend,\hgtt-2-0.33){non-tight bounds};

\draw(0,0)circle(2pt)node[right]{$\frac{1}{4}$[CMSLX16]$^\flat$};

\draw[->](0,0.1)--node[right]{$\frac{\ximax}{2(2\ximax-1)}$[\textbf{Thm.~\ref{thm::mr}(ii)}]}(0,\hgt-0.1);
\draw(0,\hgt)circle(2pt)node[below right]{$\frac{1}{2}$[GNR14]};

\draw[->](0.1,0.3*\hgt)--node[below]{$\frac{F(\ximax)}{(1+\kmin)(e^{1/\kmin}-1)}$[\textbf{Thm.~\ref{thm::mr}(i)}]}(\wdt-0.1,0.3*\hgt);
\draw[->](0.1,\hgt)--node[above]{$\frac{1-1/e}{(1+\kmin)(1-e^{-1/\kmin})}$[\textbf{Thm.~\ref{thm::mr}(iii)}]}(\wdt-0.1,\hgt);

\filldraw(\wdt,0)circle(2pt)node[above right]{$1-\frac{1}{\sqrt{e}}\approx.393$};
\draw[->](\wdt,0.1)--node[right,align=left]{$F(\ximax)$\\$[$\textbf{Cor.~\ref{cor::kToInfty}},\textbf{Thm.~\ref{thm::ub}}$]$}(\wdt,\hgt-0.1)[ultra thick];
\filldraw(\wdt,\hgt)circle(2pt)node[below right,align=left]{$1-\frac{1}{e}\approx.632$\\$[$GNR14$]$};

\filldraw(\wdt+\wdtt,0)circle(2pt)node[above right]{$1-\frac{1}{\sqrt{e}}$};
\node[\zAxisColor] at (\wdt+\wdtt,-\axisLabelDist){(any $\kmin$)};
\draw(\wdt+\wdtt,-1.5*\axisLabelDist)node[\zAxisColor,below]{\ranking in Deterministic Case};
\draw[->](\wdt+\wdtt,0.1)--node[right,align=left]{$F(\ximax)$\\$[$\textbf{Thm.~\ref{thm::djk}},\textbf{Thm.~\ref{thm::ub}}$]$}(\wdt+\wdtt,\hgt-0.1)[ultra thick];
\filldraw(\wdt+\wdtt,\hgt)circle(2pt)node[below right,align=left]{$1-\frac{1}{e}$\\$[$AGKM11$]$};

\end{tikzpicture}
\end{center}
{\footnotesize $\flat$ The smallest guarantee of $\frac{1}{4}$ in this diagram is also implied by the results of \cite{CMSLX16}.}
\caption{Competitive ratios guarantees when each item has two prices.  The guarantees increase from bottom to top (as the maximum ratio of an item's high to low price, $\ximax$, ranges from $\infty$ to 1), and from left to right (as the minimum starting inventory, $\kmin$, ranges from 1 to $\infty$).}
\label{fig::cube}
\end{figure}

If every item has two prices and $\ximax$ denotes the maximum ratio of an item's high price to low price,
then $\min_iF(\cP_i)$ equals $F(\ximax)$
because $F$ is decreasing.
Also letting $\kmin$ denote the minimum starting inventory among the items, we plot, in Figure~\ref{fig::cube}, our competitive ratio guarantees for \balance and \ranking as both $\ximax$ and $\kmin$ range over $[1,\infty]$.
This guarantee equals $F(\ximax)$ in the asymptotic regime or deterministic case, which is tight.

The lower bound on the competitive ratio guarantee when each item has at most two prices occurs as $\ximax\to\infty$, in which case $F(\ximax)=1-\frac{1}{\sqrt{e}}\approx0.393$.
This is greater than the naive bound of $\frac{1}{2}(1-\frac{1}{e})$,
which would arise from randomly choosing between 2 prices and then using a $(1-\frac{1}{e})$-competitive algorithm on the chosen prices.
Thus, using a function like $\Phi$ to integrate the selection of prices with the allocation across items is necessary for achieving the optimal competitive ratio.

Our
guarantees may not be tight in the non-asymptotic, non-deterministic setting, which is an important open problem even in the single-price case \citep{DJK13}.
Nonetheless, as $\kmin$ increases, our bounds sharply approach the tight guarantee from the asymptotic regime.
In the single-price case, our bound is a factor of $(1+\kmin)(1-e^{-1/\kmin})$ from the tight guarantee of $1-1/e$, improving the previous-best-known dependence on $\kmin$ from \citet{GNR14}.

\subsection{Simulations on Hotel Data Set of \cite{BFG09}}

We first summarize the general benefits of applying the algorithms from competitive analysis.  In contrast to traditional algorithms, which optimize based on a forecast of future demand, or attempt to learn the demand, competitive algorithms guarantee some performance ratio under the worst case, and operate without any demand information.  Most immediately, they are useful for products with highly unpredictable demand \citep{BQ09,LGBK08}, or for initializing new products with no historical sales data \citep{vRM00}.  Second, by eschewing stochastic processes for generating demand, competitive algorithms are usually simple and flexible, leading to clean insights about the problem \citep{BEY05}.  Third, past research has reported on cases where competitive algorithms perform well in practice \citep{FHKMS10}, or on average in numerical experiments \citep{GNR14,CMSLX16}.

In Section~\ref{sect::simul}, we run simulations on the publicly-accessible hotel data set of \citet{BFG09}.  We use the product availability information to estimate customer choice models, and use the sequence of transactions as the sequence of arrivals.  This leads to an online assortment problem like in \cite{GNR14}, but with multiple prices (advance-purchase rate, rack rate, etc.) for each item (King room, Two-double room, etc.).  We compare the performance of our \balance algorithm to various benchmarks and forecasting algorithms.

The main conclusion from our simulations is that the best performance is achieved by \textit{hybrid} algorithms (see \cite{GNR14}).  These are forecasting-based algorithms which continuously reference our forecast-independent value functions $\Phi_1,\ldots,\Phi_n$, and adjust their decisions accordingly.  Although this only changes a small fraction ($\approx5\%$) of decisions, these tend to be the decisions where the forecast is being most overconfident.  Therefore, not only does this boost average performance, it drastically reduces the variance in performance caused when the forecast is wrong.

\subsection{Other Related Work}

We briefly mention some papers which study online resource allocation problems under other arrival models or performance metrics.

When a stochastic process generating the arrivals is given as input, the resulting optimization problem is generally still computationally intractable.  Nonetheless, many effective heuristics have been proposed under various models of online resource allocation \citep{ZC05,JK12,CF12,CF13}.  These heuristics can earn $\frac{1}{2}$ of the LP optimum in general settings \citep{CF09,WTB15}.  \cite{MGS12} derive an improved performance ratio when the stochastic process is IID.

Competitive/approximation ratios both analyze the \textit{fraction} of optimum achieved by an algorithm, but online resource allocation problems are also often analyzed under the \textit{regret} metric, which measures the \textit{difference} from optimum.
This work often focuses on learning some unknown underlying stochastic model \citep{BKS13,FSLW16}.
On the other hand, queueing-theoretic analyses have also been performed given a known stochastic model \citep{RW08}.
Unlike in competitive analysis, all of these papers tend to focus on asymptotic performance as the number of customers grows to infinity.
Finally, a recent metric which has been introduced is \textit{regret ratio} \citep{ZSQH16}.  For a comprehensive review of different metrics under different models of demand (for a single item), we refer to \cite{AC11}.

\subsection{Organization of Paper}

Throughout Sections~\ref{sect::2}--\ref{sect::countereg} of this paper, we analyze a simplified model where each customer is \textit{offered a single item at a single price} (but her purchase decision is still stochastic).
This avoids the complexities of assortment optimization while still capturing our main techniques.
In Section~\ref{sect::generalizations}, we discuss the generalizations to the assortment and Adwords settings.
In Section~\ref{sect::simul}, we display the results of our simulations on the hotel data set.

\section{Problem Definition, Algorithm Sketch, and Theorem Statements}\label{sect::2}

A firm is selling $n\in\bN$ different items.  Each item $i\in[n]$\footnote{For a general positive integer $b$, let $[b]$ denote the set $\{1,\ldots,b\}$.} starts with a fixed inventory of $k_i\in\bN$ units, and could be offered at any price in its price set $\cP_i$.
Throughout most of this paper, we assume that each $\cP_i$ consists of $m_i\in\bN$ discrete prices satisfying $0<r^{(1)}_i<\ldots<r^{(m_i)}_i$.
We will refer to $r^{(j)}_i$ as ``price $j$ of item $i$'', and define 
$r^{(0)}_i:=0$.
We extend to the case where $\cP_i$ is a continuum of prices
in Appendix~\ref{appx::continuum}.

There are $T\in\bN$ customers arriving sequentially.  Upon the arrival of customer $t\in[T]$, the firm is given $p^{(j)}_{t,i}$, the probability\footnote{
If $\cP_i$ is a continuum of prices, then we need to assume that the purchase probabilities can be input compactly.  There are many parametric models for doing so, e.g.\ linear demand, where the purchase probability is $a-bP$ for prices $P$ lying in an interval $[\Rmin,\Rmax]$.
}
that customer $t$ would buy item $i$ at price $j$, for all $i\in[n]$ and $j\in[m_i]$.\footnote{
These probabilities can be 0 for items the customer is not interested in, or prices that are too high.
}  The firm chooses up to one of the items $i$ with remaining inventory and offers it to customer $t$, at any price $j\in[m_i]$.  The customer accepts the offer with probability $p^{(j)}_{t,i}$, in which case the firm earns revenue $r^{(j)}_i$, and the inventory of item $i$ is decremented by 1.

We divide the elements defined above into:
\begin{enumerate}
\item The \textit{Setup} $\cS$, consisting of parameters known at the start: $\big(n,(k_i,m_i,r^{(1)}_i,\ldots,r^{(m_i)}_i)_{i\in[n]}\big)$; and
\item The \textit{Arrival sequence} $\cA$, consisting of parameters revealed over time: $\big(T,(p^{(j)}_{t,i})_{t\in[T],i\in[n],j\in[m_i]}\big)$.
\end{enumerate}

An \textit{online algorithm} must decide, on any setup $\cS$, what to offer to each customer $t$.
This decision can be based on only the setup $\cS$, the past arrivals/purchase realizations, and the purchase probabilities $p^{(j)}_{t,i}$ of the present customer $t$; the online algorithm does not know the purchase probabilities associated with future customers.
For an online algorithm, let $\ALG(\cS,\cA)$ denote the revenue earned on a run on setup $\cS$ with arrival sequence $\cA$, which is a random variable with respect to the customers' purchase decisions as well, as any randomness in the algorithm's decisions.

Meanwhile, we can write the following LP for setup $\cS$ with arrival sequence $\cA$:
\begin{subequations}\label{primal}
\begin{align}
\max\sum_{t=1}^T\sum_{i=1}^n\sum_{j=1}^{m_i}p^{(j)}_{t,i}r^{(j)}_ix^{(j)}_{t,i} & & \label{primal::obj} \\
\sum_{t=1}^T\sum_{j=1}^{m_i}p^{(j)}_{t,i}x^{(j)}_{t,i} &\le k_i &i\in[n] \label{primal::inv} \\
\sum_{i=1}^n\sum_{j=1}^{m_i}x^{(j)}_{t,i} &\le1 &t\in[T] \label{primal::offer} \\
x^{(j)}_{t,i} &\ge0 &t\in[T],i\in[n],j\in[m_i] \label{primal::nonneg}
\end{align}
\end{subequations}
LP (\ref{primal}) encapsulates the execution of any algorithm, which could make full use of the arrival sequence $\cA$ at the start, on setup $\cS$---$x^{(j)}_{t,i}$ represents the unconditional probability of the algorithm offering item $i$ at price $j$ to customer $t$; (\ref{primal::inv}) enforces that starting inventories are respected; (\ref{primal::offer}) enforces that at most one combination of item and price is offered to each customer; and objective function (\ref{primal::obj}) represents the expected revenue earned by the algorithm.
Let $\OPT(\cS,\cA)$ denote its optimal objective value.
Note that although $\OPT(\cS,\cA)$ knows the arrival sequence in advance, it does not know the outcomes of the customers' potential purchase decisions.

For a fixed online algorithm and any setup $\cS$, the online algorithm is said to \textit{achieve a competitive ratio of} $c$ on $\cS$, if
\begin{align} \label{eqn::CR}
\frac{\bE[\ALG(\cS,\cA)]}{\OPT(\cS,\cA)}\ge c\text{\ \ \ for all arrival sequences }\cA.
\end{align}
In this paper, we will allow the competitive ratio guarantee $c$ to depend on parameters in the setup $\cS$,
and derive results that hold for any $\cS$.

Definition (\ref{eqn::CR}) provides a guarantee on $\bE[\ALG(\cS,\cA)]$ relative to any algorithm which could have been possible, due to the following standard result.

\begin{lemma}\label{lem::LPbound}
$\OPT(\cS,\cA)$ is an upper bound on the expected revenue of any algorithm, which could make full use of the arrival information at the start, on setup $\cS$ with arrival sequence $\cA$.
\end{lemma}

The proof of Lemma~\ref{lem::LPbound} is deferred to Appendix~\ref{appx::first}.  The definition of $\OPT(\cS,\cA)$ based on the LP is standard in problems with both stochastic purchase realizations and arbitrary customer arrivals---we refer to \citet{MP12,GNR14} for its justification.

In the \textit{deterministic} case of our problem, every $p^{(j)}_{t,i}$ is 0 or 1.  The problem can be simplified by letting $j_{t,i}=\max\{j\in[m_i]:p^{(j)}_{t,i}=1\}$, with $j_{t,i}=0$ if the set is empty, for all $t\in[T]$ and $i\in[n]$.  We say that item $i$ is \textit{assigned} to customer $t$ to indicate that $i$ is offered to customer $t$ at price $j_{t,i}$, which results in a sale; there is no reason to offer any other price.  Customer $t$ can also be \textit{rejected}, e.g. if $j_{t,i}$ is low for every $i$.  In the deterministic case, the LP (\ref{primal}) is integral, so $\OPT(\cS,\cA)$ is \textit{equal} to the revenue of the best algorithm knowing the arrival sequence at the start.

\subsection{Construction of Value Function for a Price Set}\label{subsect::phi}

In this section, we specify a value function $\Phi_{\cP}$ and a number $F(\cP)$, for any price set $\cP$ consisting of $m$ discrete prices with $0<r^{(1)}<\ldots<r^{(m)}$.
The derivation of $\Phi_{\cP}$ and $F(\cP)$, as well as the case where $\cP$ is a continuum of prices, are deferred to Appendix~\ref{appx::deriving}.

Consider an item with price set $\cP$.
Following the description from Section~\ref{subsect::introBidPrice}, we will interpret $\Phi_{\cP}$ to be a function of $w\in[0,1]$, which is the fraction of the item's starting inventory which has been sold.
$\Phi_{\cP}(w)$ specifies the value that should be placed on one unit of the item's inventory, when its fraction sold is $w$.

First we define ``booking limits'' $\alpha^{(1)},\ldots,\alpha^{(m)}$, which are the fractions of starting inventory ``reserved'' for the respective fares $r^{(1)},\ldots,r^{(m)}$, via the following proposition.
\begin{proposition}\label{prop::solveSys} Let $r^{(1)},\ldots,r^{(m)}$ be any numbers satisfying $0<r^{(1)}<\ldots<r^{(m)}$.  Then there is a unique set of positive values $\alpha^{(1)},\ldots,\alpha^{(m)}$ which sum to 1 and satisfy
\begin{equation}\label{eqn::solveAlpha}
1-e^{-\alpha^{(1)}}=\frac{1}{1-r^{(1)}/r^{(2)}}\cdot(1-e^{-\alpha^{(2)}})=\ldots=\frac{1}{1-r^{(m-1)}/r^{(m)}}\cdot(1-e^{-\alpha^{(m)}}).
\end{equation}
There is also a different, unique set of positive values $\sigma^{(1)},\ldots,\sigma^{(m)}$ which sum to 1 and satisfy
\begin{equation}\label{eqn::solveSigma}
\sigma^{(1)}=\frac{1}{1-r^{(1)}/r^{(2)}}\cdot\sigma^{(2)}=\ldots=\frac{1}{1-r^{(m-1)}/r^{(m)}}\cdot\sigma^{(m)}.
\end{equation}

\end{proposition}
The proof of Proposition~\ref{prop::solveSys} is elementary and deferred to Appendix~\ref{appx::first}.  While finding the exact solution to (\ref{eqn::solveAlpha}) requires finding the roots of a degree-$m$ polynomial, a numerical solution can easily be found via bisection search.

Proposition~\ref{prop::solveSys} contrasts $\alpha^{(1)},\ldots,\alpha^{(m)}$ in (\ref{eqn::solveAlpha}) with the booking limits $\sigma^{(1)},\ldots,\sigma^{(m)}$ in (\ref{eqn::solveSigma}) originally derived by \cite{BQ09}, which are optimal
for selling a \textit{single} item with price set $\{r^{(1)},\ldots,r^{(m)}\}$.
With $\alpha^{(1)},\ldots,\alpha^{(m)}$, we can now complete the definition of $\Phi_{\cP}$.
\begin{definition}\label{defn::phi}
Define the following, based on the values of $\alpha^{(1)},\ldots,\alpha^{(m)}$ from Proposition~\ref{prop::solveSys}:
\begin{itemize}
\item $L^{(j)}$: the sum $\sum_{j'=1}^j\alpha^{(j')}$, defined for all $j=0,\ldots,m$ (note that $L^{(0)}=0$ and $L^{(m)}=1$);
\item $\ell(\cdot)$: a function on $[0,1]$, where $\ell(w)$ is the unique $j\in[m]$ for which $w\in[L^{(j-1)},L^{(j)})$ (note that $\ell(L^{(j)})=j+1$ for $j=0,\ldots,m-1$; we define $\ell(L^{(m)})$ to be $m$).
\end{itemize}
The \textit{value function} $\Phi_{\cP}$ for price set $\cP$ is then defined over $w\in[0,1]$ by:
\begin{equation}\label{eqn::phi}
\Phi_{\cP}(w)=r^{(\ell(w)-1)}+(r^{(\ell(w))}-r^{(\ell(w)-1)})\frac{\exp(w-L^{(\ell(w)-1)})-1}{\exp(\alpha^{(\ell(w))})-1}.
\end{equation}
\end{definition}

An example of $\Phi_{\cP}$ for $\cP=\{150,450\}$ was plotted in the Introduction, in Figure~\ref{fig::valueFn}.
In general, $\Phi_{\cP}$ is continuously increasing and piecewise-convex over $m$ \textit{segments} of lengths $\alpha^{(1)},\ldots,\alpha^{(m)}$, separated by \textit{segment borders} $L^{(0)},\ldots,L^{(m)}$.  For each $j$, $\Phi$ reaches the value of $r^{(j)}$ at $L^{(j)}$.

\begin{definition}\label{defn::FG}
For price set $\cP=\{r^{(1)},\ldots,r^{(m)}\}$, let $F(\cP)=1-e^{-\alpha^{(1)}}$ and $G(\cP)=\sigma^{(1)}$, where $\alpha^{(1)}$ and $\sigma^{(1)}$ are the values from Proposition~\ref{prop::solveSys}.
\end{definition}

Our competitive ratio guarantees will be based on the functions $F$ and $G$.
It can be checked that $F$ maps a price set $\cP$ to $[1-e^{-1/m},1-e^{-1}]$ and $G$ maps the price set to $[1/m,1]$, where $m$ is the number of prices in $\cP$.
When $m=1$, our value function is $\Phi_{\cP}(w)=r^{(1)}\cdot\frac{e^{w}-1}{e-1}$, which can be related back to the existing multiplicative ``penalty functions'' from the single-price case.

\subsection{Sketch of our MULTI-PRICE BALANCE and MULTI-PRICE RANKING Algorithms}\label{subsect::algSketch}

Having defined the value function $\Phi_{\cP}$ for an arbitrary price set $\cP$, we now sketch our algorithms.

We start with \ranking, which is simpler.
It assumes that $k_i=1$ for all $i$, which does not lose generality since an item that starts with multiple units of inventory can be transformed into multiple disparate items.  At the start, the algorithm fixes for each item $i$ a random seed $W_i$, drawn independently and uniformly from $[0,1]$.  It then treats $\Phi_{\cP_i}(W_i)$ as the bid price for the single unit of item $i$: it offers to each customer $t$ the available item $i$ and price $j$ maximizing the expected pseudorevenue, $p^{(j)}_{t,i}\big(r^{(j)}_i-\Phi_{\cP_i}(W_i)\big)$.

\ranking hedges against the ambiguity in customer arrivals using randomness, which is standard in competitive analysis.  The random seed $W_i$ determines the random minimum price at which the algorithm is willing to sell item $i$, as well as a random priority for selling $i$ when the algorithm is choosing between multiple items.

We now sketch \balance, which updates the bid price of each item $i$ based on the fraction $w_i$ of its $k_i$ units which has been sold.
However, the algorithm does not directly use $\Phi_{\cP_i}(w_i)$ as the bid price of item $i$, because $w_i$ would always be a multiple of $\frac{1}{k_i}$, while the booking limits and segment borders which $\Phi_{\cP_i}$ is based on may not be multiples of $\frac{1}{k_i}$.
Instead, the algorithm first uses a \textit{randomized procedure} for rounding the booking limits in $\Phi_{\cP_i}$ to multiples of $\frac{1}{k_i}$.

Specifically, at the start, the algorithm fixes for each item $i$ \textit{random} segment borders $\tL^{(0)}_i,\ldots,\tL^{(m_i)}_i$, which are multiples of $\frac{1}{k_i}$ satisfying $0=\tL^{(0)}_i\le\ldots\le\tL^{(m_i)}_i=1$.
We note that having $\tL^{(i)}=\tL^{(i-1)}$ is possible (and guaranteed to happen if $m_i>k_i$), in which case the $i$'th segment has length zero.
In either case, the realizations of $\tL^{(0)}_i,\ldots,\tL^{(m_i)}_i$
imply a random value function $\tPhi_i$ for item $i$, which is a \textit{perturbation} of $\Phi_{\cP_i}$.
Function $\tPhi_i$ is defined on $\{0,\frac{1}{k_i},\ldots,1\}$, since the fraction sold $w_i$ is always a multiple of $\frac{1}{k_i}$, and also satisfies $0=\tPhi_i(0)\le\tPhi_i(\frac{1}{k_i})\le\ldots\le\tPhi_i(1)$.
At any point in time, \balance treats $\tPhi_i(w_i)$ as the bid price for item $i$: it offers to each customer $t$ the item $i$ and price $j$ maximizing
\begin{equation}\label{eqn::clarificationCarriVAT}
p^{(j)}_{t,i}\big(\tPhi_i(\tL^{(j)}_i)-\tPhi_i(w_i)\big).
\end{equation}
In (\ref{eqn::clarificationCarriVAT}), the definition of pseudorevenue at price $j$ is $\tPhi_i(\tL^{(j)}_i)-\tPhi_i(w_i)$ instead of $r^{(j)}_i-\tPhi_i(w_i)$.
This is because we want the expected pseudorevenue to be 0 when $w_i=\tL^{(j)}_i$.
In general, the realized $\tPhi_i$ will be close to $\Phi_{\cP_i}$, so that $\tPhi_i(\tL^{(j)}_i)\approx\Phi_{\cP_i}(L^{(j)}_i)=r^{(j)}_i$.  In the asymptotic regime with $k_i\to\infty$, $\tPhi_i$ is deterministically initialized to $\Phi_{\cP_i}$.  However, for small $k_i$, optimizing a \textit{randomized procedure} for initializing $\tPhi_i$ (based on $r^{(1)}_i,\ldots,r^{(m_i)}_i$ as well as $k_i$) instead of having the deterministic $\Phi_{\cP_i}$ (which is based on only $r^{(1)}_i,\ldots,r^{(m_i)}_i$) allows
us
to achieve a greater competitive ratio.

\subsection{Statements of Our Results}\label{subsect::stmt}

Our competitive ratio results are based on the universal functions $F$ and $G$ from Definition~\ref{defn::FG}, which assign a number to every price set $\cP$.

\begin{theorem}\label{thm::mr}
For any setup, with price sets denoted by $\cP_1,\ldots,\cP_n$ and starting inventories denoted by $k_1,\ldots,k_n$, \balance achieves a competitive ratio of $\min_i\tCR_i$, where for each item $i$, $\tCR_i$ is lower-bounded by all of: (i) $\frac{F(\cP_i)}{(1+k_i)(e^{1/k_i}-1)}$; (ii) $\frac{G(\cP_i)}{2}$; and (iii) $\frac{1-1/e}{(1+k_i)(1-e^{-1/k_i})}$ if $|\cP_i|=1$.
\end{theorem}

\begin{corollary}\label{cor::kToInfty}
\balance achieves a competitive ratio approaching $\min_iF(\cP_i)$ as each starting inventory $k_i$ approaches $\infty$.
\end{corollary}

\begin{corollary}\label{cor::ozan}
Suppose that each item has at most $m$ discrete prices and at least $k$ units of starting inventory.  Then the competitive ratio achieved by \balance is lower-bounded by $\frac{1-e^{-1/m}}{(1+k)(e^{1/k}-1)}$, which approaches $1-e^{-1/m}$ as $k$ approaches $\infty$.
\end{corollary}

Theorem~\ref{thm::mr} is our general result, where for each $i$, $\tCR_i$ is determined by the randomized procedure used to initialize the value function $\tPhi_i$ for item $i$.

Lower bound (i) on $\tCR_i$ is attained by a randomized procedure which perturbs the ``ideal'' value function $\Phi_{\cP_i}$ to define $\tPhi_i$.
This perturbation loses a factor of $(1+k_i)(e^{1/k_i}-1)$ in the denominator, which decreases to 1 as $k_i\to\infty$, resulting in Corollary~\ref{cor::kToInfty}.
Corollary~\ref{cor::ozan} is a further simplification of the bound presented, using the fact that $F(\cP)\ge1-e^{-1/|\cP|}$.
Meanwhile, lower bound (ii) is attained by solving an optimization problem for the best randomized procedure to define $\tPhi_i$ when $k_i=1$; this procedure is not based on perturbing $\Phi_{\cP_i}$ and the bound is based on 
$G(\cP_i)$ instead of $F(\cP_i)$.
Finally, lower bound (iii) is an improvement of (i) in the single-price case, where we have gained a factor of $e^{1/k_i}$ in the denominator.  It simplifies and improves the dependence on $k_i$ from \cite{GNR14}.

\balance is formalized and Theorem~\ref{thm::mr} is proven in Section~\ref{sect::balance}.  We explain the ideas behind our primal-dual analysis, why we need \textit{random} value functions, and how to overcome the ensuing analytical challenges.

\begin{theorem}\label{thm::djk}
For any setup in the deterministic case, \ranking achieves a competitive ratio of $\min_iF(\cP_i)$.
\end{theorem}

\ranking is formalized and Theorem~\ref{thm::djk} is proven in Section~\ref{sect::ranking}.
Our analysis builds upon the framework of \cite{DJK13} and extends it to handle multiple prices.  

\begin{theorem}\label{thm::ub}
Let $k$ be any positive integer.
Let $\cP$ be any price set consisting of $m$ prices with $0<r^{(1)}<\ldots<r^{(m)}$.
Then there exists a setup $\cS$ where each item has starting inventory $k$ and price set $\cP$, along with a distribution over arrival sequences $\cA$ falling in the deterministic case, for which no online algorithm can have expected revenue greater than $F(\cP)\cdot\bE_{\cA}[\OPT(\cS,\cA)]$.
\end{theorem}

Theorem~\ref{thm::ub} is proven in Section~\ref{sect::countereg}.
Since the starting inventory $k$ can be made arbitrarily large and the arrival sequences fall in the deterministic case, Theorem~\ref{thm::ub} implies that the competitive ratio guarantees in Corollary~\ref{cor::kToInfty} and Theorem~\ref{thm::djk} are tight, via Yao's minimax principle \citep{Yao77}.

Our counterexample is based on those from \citet{KVV90,MSVV07,GNR14}, where 
a large number of customers arrive according to a random permutation chosen uniformly from all possible permutations.
In our case however, the customers are further split into $m$ ``phases'', where the customers in phase $j$ are willing to pay $r^{(j)}$ for any of the items they are interested in.  The phases lengths are optimized by an adversary to minimize the competitive ratio.  

Interestingly, on the existing counterexamples, the random permutation implies that all (reasonable) algorithms are indifferent and have the same performance.
By contrast, on our counterexample with the adversarially-optimized phase lengths, there is a \textit{unique} optimal algorithm given the distribution over arrival sequences $\cA$.
When $k\to\infty$, this unique algorithms turns out to be our \balance and \ranking algorithms, which coalesce to the same algorithm in the asymptotic regime.
This coalescence phenomenon has been noted in the single-price case as well by \citet{AGKM11}.

\begin{proposition}\label{prop::comparison}
For $m\ge2$ prices satisfying $0<r^{(1)}<\ldots<r^{(m)}$, from which $\alpha^{(1)}$ and $\sigma^{(1)}$ are defined according to Proposition~\ref{prop::solveSys}, the following inequalities hold:
\begin{gather}
(1-\frac{1}{e})\cdot\sigma^{(1)}<1-e^{-\sigma^{(1)}}<1-e^{-\alpha^{(1)}}; \label{eqn::naiveSubopt} \\
\frac{1}{1+\ln\frac{r^{(m)}}{r^{(1)}}}<\sigma^{(1)}; \label{eqn::BQContinuum} \\
1-e^{-\alpha}<1-e^{-\alpha^{(1)}},\text{ where }\alpha\text{ is the unique solution to }1-e^{-\alpha}=\frac{1-\alpha}{\ln\frac{r^{(m)}}{r^{(1)}}}. \label{eqn::usContinuum}
\end{gather}
\end{proposition}

Finally, Proposition~\ref{prop::comparison}, which is proven in Appendix~\ref{appx::first}, puts our tight competitive ratio of $1-e^{-\alpha^{(1)}}$ into perspective.  $\sigma^{(1)}$ is the existing tight competitive ratio for a single item, while $1-\frac{1}{e}$ is the existing tight competitive ratio for multiple items with one price each.  (\ref{eqn::naiveSubopt}) shows that our competitive ratio for multiple items with multiple prices is not a naive combination of the existing competitive ratios; our algorithms also cannot be obtained by naively combining existing algorithms.

With a single item whose price can take any value in the continuum $[r^{(1)},r^{(m)}]$, the tight competitive ratio is $\frac{1}{1+\ln(r^{(m)}/r^{(1)})}$ \citep{BQ09}.  
(\ref{eqn::BQContinuum}) says that if the prices are restricted to a discrete subset of $[r^{(1)},r^{(m)}]$, then the competitive ratio of $\sigma^{(1)}$ can only be larger.

We have a corresponding relationship in the multi-price setting.
$1-e^{-\alpha}$, with $\alpha$ as defined\footnote{
$\alpha$ can be solved to equal $1-W(Re^{R-1})/R$, where $W$ is the inverse of the function $f(x)=xe^x$, and $R=\ln(r^{(m)}/r^{(1)})$---see Appendix~\ref{appx::continuum}.
} in (\ref{eqn::usContinuum}), is our competitive ratio when there are multiple items whose price sets are $[r^{(1)},r^{(m)}]$.
(\ref{eqn::usContinuum}) says that if the prices are restricted to a discrete subset of $[r^{(1)},r^{(m)}]$, then the competitive ratio of $1-e^{-\alpha^{(1)}}$ can only be larger.

\section{MULTI-PRICE BALANCE and the Proof of Theorem~\ref{thm::mr}}\label{sect::balance}

\balance, as sketched in Subsection~\ref{subsect::algSketch}, is formalized in Algorithm~\ref{alg::balance}.  For now, we consider a generic randomized procedure for initializing $\tL^{(0)}_i,\ldots,\tL^{(m_i)}_i$ and $\tPhi_i$ in Step~\ref{line::init}, where the realized initializations always satisfy the following monotonicity conditions:
\begin{gather}
\tL^{(0)}_i,\ldots,\tL^{(m_i)}_i\in\{0,\frac{1}{k_i},\ldots,1\},\ 0=\tL^{(0)}_i\le\ldots\le\tL^{(m_i)}_i=1; \label{eqn::random_segment_borders} \\
\tPhi_i(0),\tPhi_i(\frac{1}{k_i}),\ldots,\tPhi_i(1)\in\bR,\ 0=\tPhi_i(0)\le\tPhi_i(\frac{1}{k_i})\le\ldots\le\tPhi_i(1). \label{eqn::random_value_fn}
\end{gather}
Since $\tPhi_i$ is non-decreasing, the expression $\tPhi_i(\tL^{(j)}_i)-\tPhi_i(\frac{N_i}{k_i})$ in (\ref{eqn::pseudorev_balance}) is non-positive once the number sold $N_i$ reaches $k_i$.  Therefore, Algorithm~\ref{alg::balance} never tries to offer an item $i$ which has stocked out.

\SingleSpacedXI
\begin{pseudocode}[t]
	\begin{algorithmic}[1]
		\State Initialize $\tL^{(0)}_i,\ldots,\tL^{(m_i)}_i,\tPhi_i$ randomly and independently for each $i\in[n]$ \label{line::init}
		\State $N_i\leftarrow0$ for all $i\in[n]$ ($N_i$ tracks the total number of copies of item $i$ sold, at any price)
		\For{$t=1,2,\ldots$}
			\State Compute
			\begin{equation}\label{eqn::pseudorev_balance}
			\max_{i\in[n],j\in[m_i]}p^{(j)}_{t,i}(\tPhi_i(\tL^{(j)}_i)-\tPhi_i(\frac{N_i}{k_i}))
			\end{equation}
			\If{the value of (\ref{eqn::pseudorev_balance}) is strictly positive}
				\State Offer any item $i^*_t$ and price $j^*_t$ maximizing (\ref{eqn::pseudorev_balance}) to customer $t$
				\If{customer $t$ accepts (occurring with probability $p^{(j^*_t)}_{t,i^*_t}$)}
					\State $Z_t\leftarrow \tPhi_{i^*_t}(\tL^{(j^*_t)}_{i^*_t})-\tPhi_{i^*_t}(N_{i^*_t}/k_{i^*_t})$ (this is the pseudorevenue earned) \label{line::Zt}
					\State $N_{i^*_t}\leftarrow N_{i^*_t}+1$ \label{line::Ni}
				\EndIf
			\EndIf
		\EndFor
	\end{algorithmic}
	\caption{\balance}\label{alg::balance}
\end{pseudocode}
\DoubleSpacedXI

\begin{theorem}\label{thm::randomizedBalance}
Suppose in Line~\ref{line::init} of Algorithm~\ref{alg::balance}, for each $i\in[n]$, the segment borders $\tL^{(1)}_i,\ldots,\tL^{(m_i)}_i$ and value function $\tPhi_i$ are randomly initialized in a way such that
\begin{align}
k_i(\tPhi_i(\frac{N+1}{k_i})-\tPhi_i(\frac{N}{k_i}))+\tPhi_i(\tL^{(j)}_i)-\tPhi_i(\frac{N}{k_i}) &\le \frac{r^{(j)}_i}{c}, &j\in[m_i],N\in\{0,\ldots,\tL^{(j)}_ik_i-1\}; \label{eqn::optimality} \\
\bE[\tPhi_i(\tL^{(j)}_i)] &\ge r^{(j)}_i, &j\in[m_i]. \label{eqn::feasibility}
\end{align}
Then Algorithm~\ref{alg::balance} achieves a competitive ratio of $c$.
\end{theorem}

Theorem~\ref{thm::randomizedBalance} identifies conditions which, when satisfied by the randomized procedure for each $i$, yields a competitive ratio of $c$.  Note that (\ref{eqn::optimality}) needs to hold
for every potential initialization of $\tPhi_i$,
while (\ref{eqn::feasibility}) only needs to hold in expectation over the initializations.  We prove Theorem~\ref{thm::randomizedBalance} in Appendix~\ref{appx::mr}, but outline its proof here and provide some intuition.

First, we take the dual of the LP (\ref{primal}):
\begin{subequations}\label{dual}
\begin{align}
\min\sum_{i=1}^nk_iy_i+\sum_{t=1}^Tz_t & & \label{dual::obj} \\
p^{(j)}_{t,i}y_i+z_t &\ge p^{(j)}_{t,i}r^{(j)}_i &t\in[T],i\in[n],j\in[m_i] \label{dual::feas} \\
y_i,z_t &\ge0 &i\in[n],t\in[T] \label{dual::nonneg}
\end{align}
\end{subequations}
By weak duality, $\OPT(\cS,\cA)$ is upper-bounded by the objective value of any feasible dual solution.

During the (random) execution of Algorithm~\ref{alg::balance}, it maintains a dual variable $y_i=\tPhi_i(\frac{N_i}{k_i})$ for each $i$.  At each time $t$, only if a sale is realized, does the algorithm set $z_t$ to a non-zero value $Z_t$ (Line~\ref{line::Zt}) and increment the $y_i$-variables by incrementing $N_{i^*_t}$ (Line~\ref{line::Ni}).  We prove three claims:
\begin{enumerate}
\item During each time $t\in[T]$, the gain in the dual objective is at most some multiple $\frac{1}{c}$ of the revenue earned by the algorithm;
\item During each time $t\in[T]$, the conditional expectation of $Z_t$ over the random purchase decision of customer $t$, combined with the current value of $y_i$, make the LHS of (\ref{dual::feas}) at least $p^{(j)}_{t,i}\cdot \tPhi_i(\tL^{(j)}_i)$, for all $i\in[n]$ and $j\in[m_i]$;
\item The expectation of $\tPhi_i(\tL^{(j)}_i)$, over the random segment borders and value function initially chosen by the algorithm, is at least $r^{(j)}_i$, for all $i\in[n]$ and $j\in[m_i]$.
\end{enumerate}
Claim~1 follows from condition (\ref{eqn::optimality}), while Claim~3 follows from condition (\ref{eqn::feasibility}).  Claims~2 and 3 can be combined to show that the dual variables $y_i$ and $z_t$ maintained by the algorithm are feasible, after taking an expectation over all sample paths.

We explain the intuition behind our idea of a \textit{random} value function, and the resulting analysis.  Even for a single item, with a small starting inventory and a large ratio $r$ from its highest to lowest price, in order to achieve a constant competitive ratio which does not scale with $r$, one must use \textit{random} booking limits \citep{BQ09}.  With multiple items, our equivalent is to have the configuration of segment borders $\tL^{(0)}_i,\ldots,\tL^{(m_i)}_i$ be random, and define an arbitrary value function $\tPhi_i$ corresponding to each one.  In order to ``average'' over these configurations in the analysis, we relax dual feasibility to only hold in expectation.  The idea of feasibility in expectation has been previously seen, but in different contexts: in \cite{DJK13}, over a random seed, and in \cite{GNR14}, over a random purchase decision (similar to our Claim~2).

\subsection{Optimizing the Randomized Procedures}

Theorem~\ref{thm::randomizedBalance} reduces the problem of deriving a competitive algorithm to that of finding a randomized procedure for initializing $\tPhi_1,\ldots,\tPhi_n$ satisfying (\ref{eqn::optimality})--(\ref{eqn::feasibility}).
We can consider this problem separately for each $i$, based on $r^{(1)}_i,\ldots,r^{(m_i)}_i$ and $k_i$, and omit the subscript $i$.

A randomized procedure consists of a distribution over the all of the configurations satisfying (\ref{eqn::random_segment_borders}), and for each configuration, values for $\tPhi(\frac{1}{k}),\tPhi(\frac{2}{k}),\ldots,\tPhi(1)$ satisfying (\ref{eqn::random_value_fn}).  We would like to find a randomized procedure which satisfies (\ref{eqn::optimality})--(\ref{eqn::feasibility}) with a maximal value of $c$.  While this optimization problem is intractable in general, we can use the intuition behind the definitions of $L^{(0)},\ldots,L^{(m)}$ and $\Phi_{\cP}$ from Subsection~\ref{subsect::phi} to specify a near-optimal randomized procedure.

\begin{definition}\label{defn::tphi}
Define the following randomized procedure for initializing $\tPhi$:
\begin{enumerate}
\item Draw a random seed $W$ uniformly from $[0,1]$;
\item For each $j$, set $\tL^{(j)}=\frac{\lfloor L^{(j)}k\rfloor+1}{k}$ if $W<L^{(j)}k-\lfloor L^{(j)}k\rfloor$, and $\tL^{(j)}=\frac{\lfloor L^{(j)}k\rfloor}{k}$ otherwise; \label{step::ppc}
\item For $q\in\{0,\frac{1}{k},\ldots,1\}$, let $\tell(q)$ be the unique $j\in[m]$ such that $\tL^{(j-1)}\le q<\tL^{(j)}$ (note that $\tell(\tL^{(j)})=j+1$ for $j=0,\ldots,m-1$; we define $\tell(\tL^{(m)})$ to be $m$).
\end{enumerate}
The value function $\tPhi$ is then defined over $q\in\{0,\frac{1}{k},\ldots,1\}$ by
\begin{equation}\label{eqn::tphi}
\tPhi(q)=\sum_{j=1}^{\tell(q)-1}(r^{(j)}-r^{(j-1)})\frac{\exp(\tL^{(j)}-\tL^{(j-1)})-1}{\exp(\alpha^{(j)})-1}+(r^{(\tell(q))}-r^{(\tell(q)-1)})\frac{\exp(q-\tL^{(\tell(q)-1)})-1}{\exp(\alpha^{(\tell(q))})-1}.
\end{equation}
\end{definition}
$\tPhi$ increases over the $m$ (possibly empty) ``segments'' of its domain $\{0,\frac{1}{k},\ldots,1\}$, which are ``bordered'' by $\tL^{(0)},\ldots,\tL^{(m)}$.  (\ref{eqn::tphi}) is similar to definition (\ref{eqn::phi}) for $\Phi_{\cP}$, except the sum in (\ref{eqn::tphi}) does not telescope, since $\tL^{(j)}-\tL^{(j-1)}$ equals $\alpha^{(j)}$ only in expectation.

Note that in Step~\ref{step::ppc} above, the random segment borders $\tL^{(0)},\ldots,\tL^{(m)}$ are rounded \textit{comonotonically} (in a perfectly positively correlated fashion) using a single seed.
This ensures that the borders are increasing as required in (\ref{eqn::random_segment_borders}), as well as the following properties.
\begin{proposition}\label{prop::ppc}
The random values of $\tL^{(0)},\ldots,\tL^{(m)}$ from Definition~\ref{defn::tphi} satisfy:
\begin{align}
\bE[\tL^{(j)}] &=L^{(j)}, &&j=0,\ldots,m; \label{eqn::ppc2} \\
|(\tL^{(j)}-\tL^{(j')})-(L^{(j)}-L^{(j')})| &\le\frac{1}{k}, &&1\le j'<j\le m. \label{eqn::ppc3}
\end{align}
\end{proposition}

(\ref{eqn::ppc3}) is the key property derived from comonotonicity: although the rounding could move each $\tL^{(j)}$ by up to $1/k$ in either direction, the distance \textit{between} two different $\tL^{(j)},\tL^{(j')}$ never changes by more than $1/k$ from $L^{(j)}-L^{(j')}$.
Proposition~\ref{prop::ppc} is then used to prove our main result about the randomized procedure from Definition~\ref{defn::tphi}.

\begin{theorem}\label{thm::largeInv}
The randomized procedure for initializing $\tPhi$ from Definition~\ref{defn::tphi} satisfies
(\ref{eqn::optimality})--(\ref{eqn::feasibility})
with $c=\frac{1-e^{-\alpha^{(1)}}}{(1+k)(e^{1/k}-1)}$.  Furthermore, if $m=1$, then the value of $c$ can be improved to $\frac{1-e^{-\alpha^{(1)}}}{(1+k)(1-e^{-1/k})}$.
\end{theorem}

Theorem~\ref{thm::largeInv} is proven in Appendix~\ref{appx::mr}.  It, in conjunction with Theorem~\ref{thm::randomizedBalance}, establishes bounds (i) and (iii) from our main result for \balance, Theorem~\ref{thm::mr}.  In Appendix~\ref{appx::mr}, we state the complete proof of Theorem~\ref{thm::mr}, including bound (ii), which involves explicitly formulating the optimization problem over randomized procedures and solving it when $k=1$.

\section{MULTI-PRICE RANKING and the Proof of Theorem~\ref{thm::djk}}\label{sect::ranking}

In Subsection~\ref{subsect::algSketch}, we sketched \ranking for our general problem.  In Algorithm~\ref{alg::ranking}, we formalize it specifically for the deterministic case, which is the case analyzed in Theorem~\ref{thm::djk}.  Note that we have assumed, without loss of generality, that $k_i=1$ for each item $i$.

\SingleSpacedXI
\begin{pseudocode}[t]
	\begin{algorithmic}[1]
		\State Initialize $W_i$ uniformly at random from $[0,1]$, independently for each $i\in[n]$
		\State $\available_i\leftarrow\textbf{true}$ for all $i\in[n]$
		\For{$t=1,2,\ldots$}
			\State Compute
			\begin{equation}\label{eqn::pseudorev_ranking}
			\max_{i\in[n],j\in[m_i]:\available_i=\textbf{true}}(r^{(j_{t,i})}_i-\Phi_{\cP_i}(W_i))
			\end{equation}
			\If{the value of (\ref{eqn::pseudorev_ranking}) is strictly positive}
				\State Offer any item $i^*_t$ maximizing (\ref{eqn::pseudorev_ranking}) to customer $t$, at price $j_{t,i^*_t}$
				\State $\available_{i^*_t}\leftarrow\textbf{false}$
			\EndIf
		\EndFor
	\end{algorithmic}
	\caption{\ranking in the Deterministic Case}\label{alg::ranking}
\end{pseudocode}
\DoubleSpacedXI

Our analysis extends the framework of \cite{DJK13} to incorporate multiple prices.  It uses the dual LP defined in (\ref{dual}), where every $p^{(j)}_{t,i}$ is 0 or 1.

If Algorithm~\ref{alg::ranking} assigns item $i$ to customer $t$ (charging price $j_{t,i}$), then we set dual variables $Z_t=r^{(j_{t,i})}_i-\Phi_{\cP_i}(W_i)$ and $Y_i=\Phi'_{\cP_i}(W_i)$, where $\Phi_{\cP_i}$ is the fixed function defined in Subsection~\ref{subsect::phi} (we ignore the measure-zero set where $\Phi'_{\cP_i}$ is undefined).
All dual variables not set during a time period are defined to be zero.  The following lemmas are proven in Appendix~\ref{appx::other}:
\begin{lemma}\label{lem::djkOpt}
If Algorithm~\ref{alg::ranking} assigns item $i$ to customer $t$, then $(1-e^{-\alpha^{(1)}_i})(Y_i+Z_t)\le r^{(j_{t,i})}_i$ w.p.1.
\end{lemma}
\begin{lemma}\label{lem::djkFeas}
Setting $y_i=\bE[Y_i],z_t=\bE[Z_t]$ for all $i,t$ forms a feasible solution to the dual LP (\ref{dual}).
\end{lemma}

The proof of Theorem~\ref{thm::djk} is then easy given these lemmas:
\proof{Proof of Theorem~\ref{thm::djk}.}
Lemma~\ref{lem::djkFeas} implies $\OPT(\cS,\cA)\le\sum_{i=1}^n\bE[Y_i]+\sum_{t=1}^T\bE[Z_t]$, via weak duality.  However, by Lemma~\ref{lem::djkOpt}, the revenue earned by Algorithm~\ref{alg::ranking} is at least $\min_{i\in[n]}\{1-e^{-\alpha^{(1)}_i}\}\cdot\big(\sum_{i=1}^nY_i+\sum_{t=1}^TZ_t\big)$, with probability 1.  Thus, $\bE[\ALG(\cS,\cA)]\ge(1-\exp(-\min_{i\in[n]}\alpha^{(1)}_i))\cdot\OPT(\cS,\cA)$.
\Halmos\endproof

\section{Randomized Counterexample and the Proof of Theorem~\ref{thm::ub}}\label{sect::countereg}

We first formalize the setup and randomized arrival sequence described in Subsection~\ref{subsect::stmt}.

There are $n\in\bN$ items, indexed by $i$, which all have $m_i=m$, $r^{(j)}_i=r^{(j)}$ for all $j$, and $k_i=k$ for some $k\in\bN$.  We think of $n$ as going to $\infty$, while $k$ is arbitrary.  Throughout this example, we often express quantities as portions $\tau$ of $n$.  We abuse notation and write $\tau n$ to refer to an integer, even if $\tau$ is irrational, since the error from rounding $\tau n$ to the nearest integer is negligible as $n\to\infty$.

The arrival sequence is randomized following the classical construction of \cite{KVV90}.  There are $T=nk$ customers, split into $n$ ``groups'' of $k$ identical customers each.  Uniformly draw a random permutation $\pi=(\pi_1,\ldots,\pi_n)$ of $(1,\ldots,n)$ from the $n!$ possibilities.  For $i\in[n]$, all $k$ customers in group $i$ would deterministically buy any item in $\{\pi_i,\ldots,\pi_n\}$.  Our construction differs from existing ones in that the $n$ groups of customers are further split into $m$ ``phases''.  Let $\beta_1,\ldots,\beta_m$ be positive numbers summing to 1, corresponding to the fraction of groups in each phase, whose values we specify later.  For all $j\in[m]$, the customers in groups $(\beta_1+\ldots+\beta_{j-1})n+1,\ldots,(\beta_1+\ldots+\beta_j)n$ are willing to pay $r^{(j)}$
for any of the items in their interest set.

\begin{definition}\label{defn::lazy}
Define the following shorthand notation for all $j=1,\ldots,m+1$:
\begin{itemize}
\item $A_j:=\sum_{\ell=j}^m\alpha^{(\ell)}$ (note that $A_1=1$ and $A_{m+1}=0$);
\item $B_j:=\sum_{\ell=j}^m\beta_{\ell}$ (note that $B_1=1$ and $B_{m+1}=0$).
\end{itemize}
\end{definition}

\begin{proposition}\label{prop::beta}
Given $m\in\bN$, $0<r^{(1)}<\ldots<r^{(m)}$, and $\alpha^{(1)},\ldots,\alpha^{(m)}$ as defined in Proposition~\ref{prop::solveSys}, there exists a unique solution to the following system of equations in variables $B_2,\ldots,B_m$:
\begin{equation}\label{eqn::propBeta}
B_mr^{(m)}e^{-\alpha^{(m)}}=\ldots=B_2r^{(2)}e^{-\alpha^{(2)}}=r^{(1)}e^{-\alpha^{(1)}},
\end{equation}
with $0<B_m<\ldots<B_2<B_1=1$.
\end{proposition}

We define $B_2,\ldots,B_m$ according to Proposition~\ref{prop::beta}.  This implies definitions for $\beta_1,\ldots,\beta_m$, which are strictly positive and sum to 1.

Now, regardless of the permutation $\pi$, the optimal algorithm allocates the $k$ copies of item $\pi_i$ to the customers in group $i$, for each $i\in[n]$, successfully serving all $T=nk$ customers and earning revenue $\sum_{j=1}^mr^{(j)}(\beta_jn)k$.  This is also the optimal objective value of the LP (\ref{primal}).  Therefore, regardless of the realized arrival sequence $\cA$, $\OPT(\cS,\cA)=\sum_{j=1}^mr^{(j)}(\beta_jn)k$ which we can rewrite as
\begin{equation}\label{eqn::ubOPT}
\sum_{j=1}^m(r^{(j)}-r^{(j-1)})B_jnk.
\end{equation}

\subsection{Upper Bound on Performance of Online Algorithms}

\begin{lemma}\label{lem::ubDominance}
The expected revenue of an online algorithm with this randomized $\cA$ is upper-bounded by the maximum value of
\begin{equation}\label{eqn::ubALG}
\sum_{j=1}^mr^{(j)}B_jn(1-e^{-\lambda_j})k
\end{equation}
subject to $0\le\lambda_j\le\ln\frac{B_j}{B_{j+1}}$ for $j\in[m-1]$, $0\le\lambda_m$, and $\sum_{j=1}^m\lambda_j\le1$.
\end{lemma}

Lemma~\ref{lem::ubDominance} drastically simplifies the analysis of the online algorithm, because it restricts to algorithms which are \textit{indifferent} to the realized permutation $\pi$, allowing for a deterministic analysis.  However, our analysis differs from existing ones
(e.g. \cite[Lem.~6]{GNR14})
in that despite the item symmetry, the online algorithm has a decision---how many customers in each phase to serve, as opposed to reserving inventory for customers in future phases.

This is controlled by the $\lambda$-variables, where $\lambda_j$ denotes the expected fraction of item $\pi_n$'s inventory sold to phase-$j$ customers.  The expected number of groups served during phase $j$ is then at most $B_jn(1-e^{-\lambda_j})$, resulting in the upper bound (\ref{eqn::ubALG}).  Constraint $\lambda_j\le\ln\frac{B_j}{B_{j+1}}$ comes from the fact that $B_jn(1-e^{-\lambda_j})$ must not exceed the total number of groups in phase $j$, $\beta_jn$.

\begin{lemma}\label{lem::valueFn}
Let $j\in[m]$ and $\tau\in[0,1]$.  The maximum value of
\begin{equation}\label{eqn::backwardInduct}
\sum_{\ell=j}^mr^{(\ell)}B_{\ell}n(1-e^{-\lambda_{\ell}})k
\end{equation}
subject to $\lambda_{\ell}\ge0$ for all $\ell=j,\ldots,m$ as well as $\sum_{\ell=j}^m\lambda_{\ell}\le\tau$ is
\begin{equation}\label{eqn::valueFn}
nk\sum_{\ell=j}^mr^{(\ell)}B_{\ell}\Big(1-\exp\big(-\alpha^{(\ell)}+\frac{A_j-\tau}{m-j+1}\big)\Big).
\end{equation}
\end{lemma}

Lemma~\ref{lem::valueFn} establishes the optimal objective value of the optimization problem from Lemma~\ref{lem::ubDominance}.  The upper bound of $\ln\frac{B_j}{B_{j+1}}$ on $\lambda_j$ for $j\in[m-1]$ turns out to not be binding.  With both lemmas, the proof of Theorem~\ref{thm::ub} is easy.

\proof{Proof of Theorem~\ref{thm::ub}.}
The value of (\ref{eqn::valueFn}) with $j=1$ and $\tau=1$ is
\begin{equation}\label{eqn::1616}
nk\sum_{\ell=1}^mr^{(\ell)}B_{\ell}(1-e^{-\alpha^{(\ell)}})=(1-e^{-\alpha^{(1)}})\sum_{\ell=1}^m(r^{(\ell)}-r^{(\ell-1)})B_{\ell}nk,
\end{equation}
where we have used (\ref{eqn::solveAlpha}) to derive the equality.  Combining Lemmas~\ref{lem::ubDominance}--\ref{lem::valueFn}, we get that the RHS of (\ref{eqn::1616}) is an upper bound on $\bE[\ALG(\cS,\cA)]$, for any online algorithm.  Meanwhile, $\OPT(\cS,\cA)$ is equal to (\ref{eqn::ubOPT}) regardless of $\cA$, which is exactly the RHS of (\ref{eqn::1616}) divided by $(1-e^{-\alpha^{(1)}})$.  We have established that $\bE[\ALG(\cS,\cA)]\le(1-e^{-\alpha^{(1)}})\bE[\OPT_{\cA}(\cS,\cA)]$, completing the proof of the theorem.
\Halmos\endproof

\begin{remark}
Suppose that $k\to\infty$.  It can be seen that our algorithm (either \balance or \ranking, which behave identically when $k\to\infty$---see \cite{AGKM11}), with booking limits $\alpha^{(1)},\ldots,\alpha^{(m)}$, is the unique optimal algorithm given this distribution over arrival sequences.  Indeed, the proof of Lemma~\ref{lem::ubDominance} shows that given $\lambda_1,\ldots,\lambda_m$, the dominant strategy for the online algorithm is to deplete the inventories of items evenly (which is possible since $k\to\infty$), in which case upper bound (\ref{eqn::ubALG}) is attained.  The proof of Lemma~\ref{lem::valueFn} shows that the unique optimal values for $\lambda_1,\ldots,\lambda_m$ are $\alpha^{(1)},\ldots,\alpha^{(m)}$.
It only remains to show that $\lambda_j=\alpha^{(j)}$ is feasible, namely $\alpha^{(j)}\le\ln\frac{B_j}{B_{j+1}}$ for $j<m$.  Applying (\ref{eqn::propBeta}), this is equivalent to showing $e^{-\alpha^{(j)}}\ge\frac{r^{(j)}e^{-\alpha^{(j)}}}{r^{(j+1)}e^{-\alpha^{(j+1)}}}$, or $e^{-\alpha^{(j+1)}}\ge\frac{r^{(j)}}{r^{(j+1)}}$, which follows from (\ref{eqn::solveAlpha}) since $1-e^{-\alpha^{(1)}}\le1$.
\end{remark}

\section{Extending our Techniques}\label{sect::generalizations}

We explain how our techniques can be extended to allow for fractional inventory consumption like in the Adwords problem \citep{MSVV07}, or offering multiple items like in the online assortment problem \citep{GNR14}.

Consider the following modification of our problem from Section~\ref{sect::2}: when customer $t$ is offered item $i$ at price $j$, she \textit{deterministically} pays $p^{(j)}_{t,i}r^{(j)}_i$ and consumes a fractional amount $p^{(j)}_{t,i}\le1$ of item $i$'s inventory, instead of paying $r^{(j)}_i$ and consuming 1 unit with probability $p^{(j)}_{t,i}$.  We assume that $\min_ik_i\to\infty$.
This generalizes the Adwords problem under the small bids assumption, by allowing each budget $i$ to be depleted at $m_i$ different rates $r^{(1)}_i,\ldots,r^{(m_i)}_i$.

For this problem, we use \balance, except since we are taking $\min_ik_i\to\infty$, we can deterministically set each $\tPhi_i=\Phi_{\cP_i}$.  The three claims used to establish Theorem~\ref{thm::randomizedBalance} are simpler: Claim~2 now holds deterministically instead of requiring a conditional expectation over $Z_t$, while Claim~3 also holds deterministically since $\tPhi_i$ is always $\Phi_{\cP_i}$.  In Theorem~\ref{thm::largeInv}, condition (\ref{eqn::optimality}) is now only satisfied under an additional error term $\varepsilon$, since $N$ is no longer a discrete integer.  Nonetheless, the rounding error $\varepsilon$ approaches 0 as $k_i\to\infty$, so the optimal competitive ratio is still achieved.

For online assortment, we use the term \textit{product} to refer to an (item, price)-combination $(i,j)$.  Consider the following modification of our problem from Section~\ref{sect::2}: upon the arrival of customer $t$, for any subset (assortment) $S$ of products and $(i,j)\in S$, we are given $p^{(j)}_{t,i}(S)$, the probability that customer $t$ would pick product $(i,j)$ when offered the choice from $S$.  After being given these probabilities, we must offer an assortment $S$ to customer $t$.  This generalizes the original online assortment problem of \citet{GNR14}, by allowing each item be offered at different prices.
We note that the assortment offered $S$ can be constrained to lie in an arbitrary downward-closed family $\cF$ of subsets of $\{(i,j):i\in[n],j\in[m_i]\}$; for example, we could disallow assortments where an item is simultaneously offered at multiple prices.
The execution of an algorithm can be encapsulated by the following modification of the LP (\ref{primal}):
\begin{subequations}\label{primalAssortment}
\begin{align}
\max\sum_{t=1}^T\sum_{S\in\cF}x_t(S)\sum_{(i,j)\in S}r^{(j)}_ip^{(j)}_{t,i}(S) & & \\
\sum_{t=1}^T\sum_{S\in\cF}x_t(S)\sum_{j:(i,j)\in S}p^{(j)}_{t,i}(S) &\le k_i &i\in[n] \label{primalAssortment::inv} \\
\sum_{S\in\cF}x_t(S) &=1 &t\in[T] \\
x_t(S) &\ge0 &t\in[T],S\in\cF
\end{align}
\end{subequations}

\balance can be directly applied to this problem, with the change that it offers the \textit{assortment} $S\in\cF$ maximizing expected pseudorevenue, $\sum_{(i,j)\in S}p^{(j)}_{t,i}(S)\big(\tPhi_i(\tL^{(j)}_i)-\tPhi_i(w_i)\big)$, to each customer $t$.  We assume the existence of an oracle for solving this single-shot assortment optimization problem, which admits an efficient algorithm under many commonly-used choice models (see \citet{CSL16} for a summary).  In the analysis, dual constraints (\ref{dual::feas}) now require $z_t\ge\sum_{(i,j)\in S}p^{(j)}_{t,i}(S)(r^{(j)}_i-y_i)$ for all $t$ and $S$, which is still implied by the conditions of Theorem~\ref{thm::randomizedBalance} as long as the choice probabilities for customers satisfy a mild \textit{substitutability} assumption (see \cite{GNR14} for details).

\section{Simulations on Hotel Data Set of \citet{BFG09}}\label{sect::simul}

We test our algorithms on the publicly-accessible hotel data set collected by \cite{BFG09}.  Based on the data, we consider a multi-price online assortment problem, as defined in Section~\ref{sect::generalizations}.

\subsection{Experimental Setup}

We consider Hotel~1 from the data set, which has more transactions than the other four hotels.  For each transaction, we use \textit{booking} to refer to the date the transaction occurred, and \textit{occupancy} to refer to the dates the customer will stay in the hotel.  We consider occupancies spanning the 5-week period from Sunday, March 11th, 2007 to Sunday, April 15th, 2007.  Although the data contains occupancies for a couple of weeks outside this range, such transactions are sparse.

We merge the different rooms into 4 categories: King rooms, Queen rooms, Suites, and Two-double rooms.  Rooms under the same category draw from the same inventory.  We merge the different fare classes into two: discounted advance-purchase fares and regular rack rates.  We use \textit{product} to refer to any of the 8 combinations formed by the 4 room categories and 2 fares.

We estimate a Multinomial Logit (MNL) choice model on these 8 products, for each of 8 customer types.  The customer types are based on the booking channel, party size, and VIP status (if any) associated with a transaction.  These types capture preference heterogeneity (for example, party sizes greater than 1 tend to prefer Suites and Two-double rooms).  The details of our choice estimation are deferred to Appendix~\ref{appx::numerical}.

We should point out that more sophisticated segmentation and estimation techniques have been employed on this data set \citep{vRV14,NFGJ14}.  Nonetheless, MNL has been reported to perform relatively well \citep[sec.~5.2]{vRV14}.  The MNL choice model is convenient for our purposes because under it, both the assortment optimization problem, as well as the choice-based LP (\ref{primalAssortment}) with exponentially many variables, can be solved efficiently \citep{TvR04,LvR08,CSL16}.

We treat each occupancy date as a separate instance of the problem, for which we define a sequence of arrivals, with one arrival for each transaction which occupies that date.  The choice probabilities for each arrival are determined by the customer type associated with the corresponding transaction.\footnote{The choice realized in that transaction was used for choice model estimation, but is not used in defining the arrival.}  The number of days in advance of occupancy that each arrival occurred is also recorded, but this information is only relevant for algorithms which attempt to forecast the remaining number of arrivals based on the remaining length of time.

Before we proceed, we discuss the limitations of our analysis and the data set:
\begin{enumerate}
\item In the data set, 55\% of the transactions occupy multiple, \textit{consecutive} days.  However, we treat such a transaction as a separate arrival in the instances for each of those occupancy dates.  While this is a simplifying assumption, the focus of our paper is on the basic allocation problem without complementarity effects across consecutive days, and our goal in using the data set is to extract an arrival pattern over time.
\item It is not possible to deduce from the data the fixed capacity for each category of room.  To compensate, we consider a wide range of starting capacities in our tests.
\item Estimating the number of customers who do not make a purchase is a standard challenge in choice modeling, which is exacerbated in this data set by the fact that the arrivals are rather non-stationary.  We test various assumptions on the weight of the no-purchase option in the MNL model for each customer type.  In general, we assume that this weight is large, which causes the revenue-maximizing assortments to be large, allowing for tension between offering large assortments which maximize immediate revenue, and offering small assortments which regulate inventory consumption (details in Appendix~\ref{appx::numerical}).
\end{enumerate}

\subsection{Instance Definition}

A test instance corresponds to a specific occupancy date, which has a finite inventory of each room category.
Each customer interested in that occupancy date arrives in sequence, after which her characteristics (channel, party size, VIP status) are revealed.  The problem is to show a \textit{personalized} assortment of (room, fare)-options to each customer.  The instances we test are described below.

\begin{itemize}
\item Arrival sequence: 35 possibilities, one for each day in the 5-week occupancy period.  We multiply the arrivals by 10 (i.e. instead of a type-1 customer followed by a type-2 customer, we have 10 type-1 customers followed by 10 type-2 customers), being interested in the high-inventory regime.  After multiplication, the average number of arrivals per day is 1340, peaking on Sundays and Mondays, although the number and breakdown of customers varies by day.
\item Number of products: 8 (room, fare)-combinations, identical for all instances.
\item Prices of products: displayed in Table~\ref{tbl::roomCategories}, identical for all instances.  These prices were determined by taking the average price of that (room, fare)-combination over all transactions.
\item Starting inventories: 3 possibilities, where we set the starting inventories to yield a desired \textit{loading factor}.
The loading factor is defined by the average (over all 35 days) number of customers per unit of starting inventory, and we use the same loading factors (1.4, 1.6, 1.8) as \citet{GNR14}.
For a fixed loading factor, all 35 instances have the same starting inventories.
The fraction of total starting inventory corresponding to each room type is based on the relative frequency with which that type is booked over all transactions (see Table~\ref{tbl::roomCategories}).

\end{itemize}
We test additional synthetic instances, where we increase the high fares and consider a greater range of loading factors, in Subsection~\ref{subsect::fareDiff}.

\begin{table}
	\TABLE
	{Details on Room Categories and Fares\label{tbl::roomCategories}}
	{\begin{tabular}{|c|c|c|c|c|}
		\hline
		\updown Room Category & Low Fare & High Fare & Fraction of Rooms \\
		\hline
		\up King & \$307 & \$361 & 52\% \\
		Queen & \$304 & \$361 & 15\% \\
		Suite & \$384 & \$496 & 13\% \\
		\down Two Double & \$306 & \$342 & 20\% \\
		\hline
	\end{tabular}}
	{}
\end{table}

\subsection{Algorithms Compared}

We compare the performances of 10 algorithms on each instance.

First we describe the forecast-independent algorithms we test.

\begin{enumerate}
\item \textbf{Myopic}: offer each customer the assortment maximizing immediate expected revenue, from the items that have not stocked out.
\item \textbf{Conservative}: only offer items at their maximum prices.\footnote{This algorithm selects between the items (at their high prices) using the algorithm of \citet{GNR14}.}
\item \textbf{GNR}: offer to each customer $t$ the assortment $S$ maximizing
\begin{equation*}
\sum_{(i,j)\in S}p^{(j)}_{t,i}(S)\cdot r^{(j)}_i\Psi(w_i),
\end{equation*}
where $w_i$ is the fraction of item $i$ sold and $\Psi$ is the inventory balancing function from \citet{GNR14}.
This would represent the algorithm of \citet{GNR14} applied to the multi-price setting (see Section~\ref{subsect::introBidPrice}).
\item \textbf{Our Algorithm}: offer to each customer $t$ the assortment $S$ maximizing
\begin{equation}\label{eqn::asstPseudorev}
\sum_{(i,j)\in S}p^{(j)}_{t,i}(S)\cdot(r^{(j)}_i-\Phi_{\cP_i}(w_i)),
\end{equation}
where $w_i$ is the fraction of item $i$ sold.
This is essentially \balance, except we have used the fixed value function $\Phi_{\cP_i}$ instead of the random value function $\tPhi_i$ to define the bid price of each item $i$, which is a simplifying approximation for the high-inventory regime.
\end{enumerate}

The Myopic and Conservative algorithms represent two extremes, where the former extracts the maximum in expectation from every customer and is optimal as the loading factor approaches 0, while the latter extracts the maximum from every unit of inventory and is optimal as the loading factor approaches $\infty$.  In-between these extremes, our algorithm attempts to balance revenue-per-customer and revenue-per-item as it selects items and prices to put in the assortment.

Next we describe the forecasting-based algorithms we test.  These algorithms all estimate the number of each type of customer yet to arrive, and then incorporate this information into the LP (\ref{primalAssortment}) to set bid prices.  They differ in how they perform the forecasting, and how frequently they update the bid prices by re-solving the LP.  Further details about these algorithms, as well as discussion of alternative algorithms, are deferred to Appendix~\ref{appx::bidPrice}.

\begin{enumerate}[resume]
\item \textbf{One-shot LP}: solve the LP only once, at the start, using the average number of customers of each type to appear on a given day.
\item \textbf{LP Resolving}: re-solve the LP every 100 arrivals, using updated forecasts and inventory counts.  During each re-solve, the estimated number of remaining customers is updated, taking into account the length of time remaining until occupancy, and the number of customers that have arrived.  The estimated type breakdown is fixed, based on the aggregate distribution.
\item \textbf{LP Learning}: same as LP Resolving, except the estimated type breakdown is also updated, based on the empirical distribution observed thus far.
\item \textbf{LP Clairvoyant}: same as LP Resolving, but given the true number of customers of each type remaining.
\end{enumerate}

Finally, we describe the hybrid algorithms we test.  These algorithms combine a forecasting algorithm with ``Our Algorithm'' as described above, based on a parameter $\gamma>1$.  For each customer $t$, the hybrid algorithm considers the expected pseudorevenue (as defined in (\ref{eqn::asstPseudorev})) of the assortment $S^{\mathsf{fcst}}$ suggested by the forecasting algorithm.  If this is at least $\frac{1}{\gamma}$ of the maximum value of (\ref{eqn::asstPseudorev}) over all assortments $S$, then the hybrid algorithm offers $S^{\mathsf{fcst}}$.  Otherwise, the hybrid algorithm offers the assortment suggested by our algorithm, which maximizes (\ref{eqn::asstPseudorev}) by definition.

\begin{enumerate}[resume]
\item \textbf{Resolve-1.5}: hybrid algorithm based on LP Resolving and parameter $\gamma=1.5$.
\item \textbf{Learn-1.5}: hybrid algorithm based on LP Learning and parameter $\gamma=1.5$.
\end{enumerate}

We selected $\gamma=1.5$ above by taking the better of the two values $1.5,2.0$ tested in \citet{GNR14}.
We did not search over $\gamma>1$ for the best $\gamma$, as the reported performance of such a hybrid algorithm would be greatly inflated, since $\gamma$ would be chosen after seeing the performance.

\subsection{Results}\label{subsect::fareSame}

On every instance, we express the performance of each algorithm as a percentage of the LP upper bound.  That is, we take the expected revenue of the algorithm (approximated over 10 runs), and divide it by the optimal objective value of the LP (\ref{primalAssortment}) with the true arrival sequence.
In Table~\ref{tbl::numericalResults}, we report the mean and standard deviation of each algorithm's percentages over the 35 arrival sequences, for each loading factor.
\begin{table}
	\scriptsize
	\TABLE
	{The percentages of optimum achieved by different algorithms.  The 3 highest percentages in each row are \textbf{bolded}.  The 3 lowest standard deviations in each row are \textit{italicized}.\label{tbl::numericalResults}}
	{\begin{tabular}{|c|c|cccc|cccc|cc|}
		\cline{1-1}\cline{3-12}
		\updown Loading & \multirow{2}{*}{} & \multicolumn{4}{c|}{Forecast-independent} & \multicolumn{4}{c|}{Forecast-dependent} & \multicolumn{2}{c|}{Hybrid} \\
		\cline{3-12}
		\updown Factor & & Myopic & Conservative & GNR & OurAlg & One-shot & Resolve & Learn & Clairvoyant & Resolve-1.5 & Learn-1.5 \\
		\hline
		\up \multirow{2}{*}{1.4} & Mean & 0.974 & 0.940 & 0.973 & \textbf{0.976} & 0.973 & 0.962 & 0.958 & 0.991 & \textbf{0.977} & \textbf{0.977} \\
		\down & Stdev & 0.023 & 0.034 & 0.020 & \textit{0.013} & \textit{0.016} & 0.039 & 0.041 & 0.008 & \textit{0.018} & 0.020 \\
		\hline
		\up \multirow{2}{*}{1.6} & Mean & 0.965 & 0.960 & 0.964 & \textbf{0.971} & 0.964 & 0.961 & 0.963 & 0.990 & \textbf{0.977} & \textbf{0.978} \\
		\down & Stdev & 0.025 & 0.036 & 0.020 & \textit{0.014} & 0.021 & 0.031 & 0.030 & 0.008 & \textit{0.008} & \textit{0.010} \\
		\hline
		\up \multirow{2}{*}{1.8} & Mean & 0.957 & \textbf{0.972} & 0.960 & 0.968 & 0.808 & 0.962 & 0.968 & 0.990 & \textbf{0.977} & \textbf{0.977} \\
		\down & Stdev & 0.020 & 0.036 & 0.017 & \textit{0.012} & 0.100 & 0.029 & 0.023 & 0.009 & \textit{0.008} & \textit{0.007} \\
		\hline
	\end{tabular}}
	{}
\end{table}

In general, our \balance algorithm is the most profitable and consistent among the forecast-independent algorithms,
while the forecast-dependent algorithms have much greater fluctuation in their performance for different occupancy days, depending on how accurate their forecasts were for that day.  LP Learning is slightly better than the others, but is most prone to overfitting in its forecasts.
We note that although the forecast-independent algorithms do not make use of information about the remaining time horizon (which can be used to estimate the remaining number of customers), they perform comparably well to the forecasting algorithms.
Nonetheless, by combining the forecasting algorithms with \balance, the hybrid algorithms are able to correct for forecast overconfidence and achieve the best performance overall (aside from the Clairvoyant algorithm, which has a perfect forecast of the future).  We find that although the hybrid algorithm only changes a small fraction ($\approx5\%$) of the forecasting algorithm's decisions, this drastically improves the profitability and consistency.

\subsection{Results under Greater Fare Differentiation}\label{subsect::fareDiff}

The instances tested in Subsection~\ref{subsect::fareSame} were ``easy'' in that there was not so much difference between selling rooms at their low or high fares.  In this subsection, we synthetically modify the higher fare for each room category to be \textit{twice} its lower fare.  We also increase the utility of the no-purchase option in the MNL model for each customer type (see Appendix~\ref{appx::numerical}), to maintain the tension between low fares which maximize expected revenue, and high fares which limit inventory consumption.

\begin{figure}[t]
\begin{center}
\includegraphics[width=\textwidth]{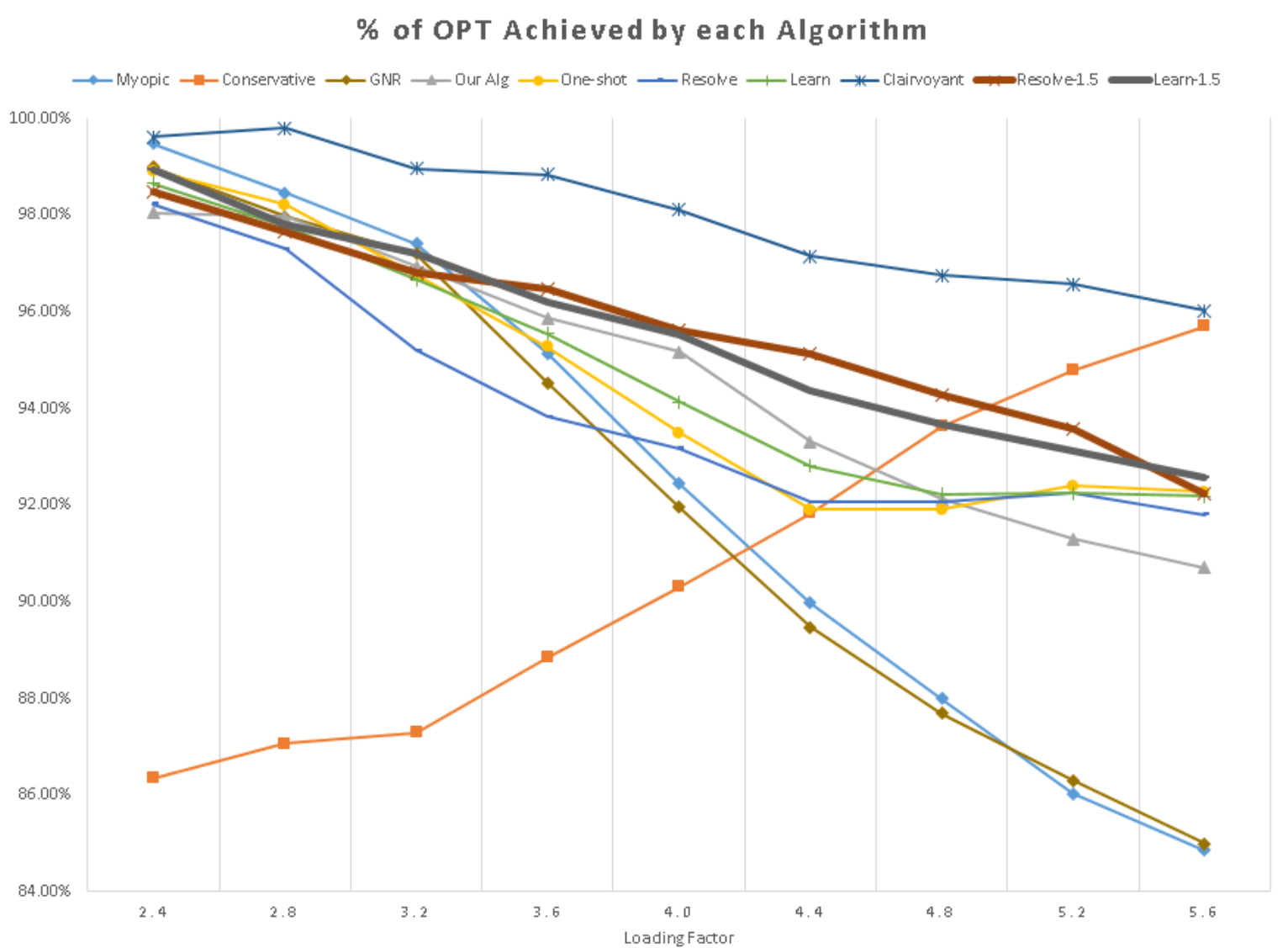}
\caption{Algorithm performances in the setting with greater fare differentiation.  The lines corresponding to the two hybrid algorithms, which perform the best overall, have been \textbf{bolded}.}\label{fig::numGraph}
\end{center}
\end{figure}

Furthermore, we test a complete range of loading factors, including both the extreme where the Myopic algorithm is optimal, and the extreme where the Conservative algorithm is optimal.  In Figure~\ref{fig::numGraph}, we plot the average percentages of optimum attained by each algorithm over the 35 arrival sequences, for each loading factor.

The conclusion again is that our two hybrid algorithms, which use forecasts but continuously reference our forecast-independent value functions, are the most profitable and consistent.
Note that our \balance algorithm comes third, and performs significantly better than the inventory-balancing algorithm of \citet{GNR14}, which is similar to the Myopic algorithm as it does not take the two different prices of the same item into account.

However, it is important to note that overall, our methodology is only relevant in scenarios where the loading factor is in-between the extremes where there is a non-trivial tradeoff.
If the loading factor is very low, and the hotel does not even get close to full on any day, then it would be best to always use the Myopic algorithm; similarly, if the loading factor is very high, and the hotel gets full every day, then it would be best to always use the Conservative algorithm.
Nonetheless, we argue that most hotels do lie in-between the extremes, where it is sometimes full and sometimes empty depending on sudden
local events.  Otherwise, the hotel either over-built or under-built in an higher-order decision.

\section{Conclusion}

Competitive analysis is a well-established methodology in sequential decision-making problems, providing a baseline decision in the absence of a reliable forecast of the future.
Previously, online resource allocation algorithms based on competitive analysis have assumed that each resource can only be converted to reward at a fixed rate.
We extend these results and derive algorithms which jointly consider the tradeoffs between different resources and different reward rates.
This broadly expands the applicability of competitive analysis in areas such as online matching, online advertising, personalized e-commerce, and appointment scheduling.

\section*{Acknowledgments}

The authors would like to thank Rong Jin of Alibaba for pointing out a detailed technical error in an earlier version of the appendix.  The authors would also like to thank Ozan Candogan and James Orlin for asking questions which led to the simpler bound presented in Corollary~\ref{cor::ozan}.
A preliminary version of this article appeared in the \textit{20th ACM Conference on Economics and Computation (EC)}, whose anonymous reviewers helped clarify the manuscript in several places.

\bibliographystyle{informs2014} 
\bibliography{bibliography} 





\clearpage

%
%
%

\begin{APPENDICES}

\section{Deferred Proofs from Section~\ref{sect::2}}\label{appx::first}

\proof{Proof of Lemma~\ref{lem::LPbound}.}
Fix any adaptive algorithm (which knows the arrival sequence, but not the realizations of the customers' purchase decisions, at the start) and consider its execution on setup $\cS$ with arrival sequence $\cA$.  Let $X^{(j)}_{t,i}$ be the indicator random variable (0 or 1) for the algorithm offering item $i$ at price $j$ to customer $t$, and $P^{(j)}_{t,i}$ be the indicator random variable for customer $t$ accepting when item $i$ is offered to her at price $j$.  On a given run, the constraints $\sum_{t=1}^T\sum_{j=1}^{m_i}P^{(j)}_{t,i}X^{(j)}_{t,i}\le k_i$ and $\sum_{i=1}^n\sum_{j=1}^{m_i}X^{(j)}_{t,i}\le1$ are satisfied.  Therefore, they are still satisfied after taking an expectation over all runs, and furthermore we can use independence to show that $\bE[P^{(j)}_{t,i}X^{(j)}_{t,i}]=\bE[P^{(j)}_{t,i}]\cdot\bE[X^{(j)}_{t,i}]=p^{(j)}_{t,i}x^{(j)}_{t,i}$.  Therefore, the algorithm must satisfy constraints (\ref{primal::inv}) and (\ref{primal::offer}) of the LP.  Since its revenue on a given run is $\sum_{t=1}^T\sum_{i=1}^n\sum_{j=1}^{m_i}P^{(j)}_{t,i}r^{(j)}_iX^{(j)}_{t,i}$, taking an expectation over it yields (\ref{primal::obj}), completing the proof.
\Halmos\endproof

\proof{Proof of Proposition~\ref{prop::solveSys}.}
The statement for $\sigma^{(j)},\ldots,\sigma^{(j)}$ is immediate from the fact that the explicit value of $\sigma^{(j)}$ is $(1-\frac{r^{(j-1)}}{r^{(j)}})(1+\sum_{j'=2}^m(1-\frac{r^{(j'-1)}}{r^{(j')}}))^{-1}$, for all $j\in[m]$.  To prove the statement for $\alpha^{(1)},\ldots,\alpha^{(m)}$, we show that the solution to the system of $n$ equations formed by (\ref{eqn::solveAlpha}) and $\alpha^{(1)}+\ldots\alpha^{(m)}=1$ is unique and strictly positive.

Let $\gamma^{(j)}=e^{-\alpha^{(j)}}$ for all $j$.  Then the constraint $\alpha^{(1)}+\ldots\alpha^{(m)}=1$ can be rewritten as $\prod_{j=1}^m\gamma^{(j)}=\frac{1}{e}$.  Furthermore, we derive from (\ref{eqn::solveAlpha}) that for all $j>1$, $\gamma^{(j)}=(1-\frac{r^{(j-1)}}{r^{(j)}})\gamma^{(1)}+\frac{r^{(j-1)}}{r^{(j)}}$.  Therefore,
\begin{equation}\label{eqn::nthRoot}
\gamma^{(1)}\cdot\prod_{j=2}^m\Big(\big(1-\frac{r^{(j-1)}}{r^{(j)}}\big)\gamma^{(1)}+\frac{r^{(j-1)}}{r^{(j)}}\Big)=\frac{1}{e}.
\end{equation}
Consider the LHS of (\ref{eqn::nthRoot}) as a function of $\gamma^{(1)}$ on $[\frac{1}{e},1]$.  This is a continuous, strictly increasing function which is at most $\frac{1}{e}$ when $\gamma^{(1)}=\frac{1}{e}$ and $1$ when $\gamma^{(1)}=1$.  Therefore, there is a unique solution with $\gamma^{(1)}\in[\frac{1}{e},1)$, and the resulting value of $\alpha^{(1)}$ is positive.  For $j>1$, since $\gamma^{(j)}$ can also be written as $\gamma^{(1)}+\frac{r^{(j-1)}}{r^{(j)}}(1-\gamma^{(1)})$, it can be seen that $\gamma^{(j)}\in[\frac{1}{e},1)$, hence the unique value for $\alpha^{(j)}$ is positive as well.
\Halmos\endproof

\proof{Proof of Proposition~\ref{prop::comparison}.}
For the first inequality in (\ref{eqn::naiveSubopt}), observe that $f(x)=\frac{x}{1-e^{-x}}$ is a strictly increasing function on $[0,1]$.  Since $\sigma^{(1)}\in(0,1)$, $\frac{\sigma^{(1)}}{1-e^{-\sigma^{(1)}}}<\frac{1}{1-\frac{1}{e}}$, which is the desired result.

For the second inequality in (\ref{eqn::naiveSubopt}), we show $\alpha^{(1)}>\sigma^{(1)}$, by showing that for all $j=2,\ldots,m$, $\alpha^{(j)}$ is a smaller multiple of $\alpha^{(1)}$ than $\sigma^{(j)}$ is of $\sigma^{(1)}$.  This suffices because both the fractions $\alpha^{(1)},\ldots,\alpha^{(m)}$ and $\sigma^{(1)},\ldots,\sigma^{(m)}$ must sum to 1.  For a given $j$, we must establish that $\frac{\alpha^{(j)}}{\alpha^{(1)}}<\frac{\sigma^{(j)}}{\sigma^{(1)}}$.  By definition, $\frac{\sigma^{(j)}}{\sigma^{(1)}}=1-\frac{r^{(j-1)}}{r^{(j)}}=\frac{1-e^{-\alpha^{(j)}}}{1-e^{-\alpha^{(1)}}}$.  Therefore, is suffices to show that $\frac{\alpha^{(j)}}{\alpha^{(1)}}<\frac{1-e^{-\alpha^{(j)}}}{1-e^{-\alpha^{(1)}}}$, or $\frac{\alpha^{(j)}}{1-e^{-\alpha^{(j)}}}<\frac{\alpha^{(1)}}{1-e^{-\alpha^{(1)}}}$.  This follows from the fact that the function $f(x)=\frac{x}{1-e^{-x}}$ is strictly increasing.

To prove (\ref{eqn::BQContinuum}), note that $\sigma^{(1)}=(1+\sum_{j=2}^m(1-\frac{r^{(j-1)}}{r^{(j)}}))^{-1}$, while $1+\ln\frac{r^{(m)}}{r^{(1)}}=1+\sum_{j=2}^m\ln\frac{r^{(j)}}{r^{(j-1)}}$.  Therefore, it suffices to show that for any $j=2,\ldots,m$, $\ln\frac{r^{(j)}}{r^{(j-1)}}>1-\frac{r^{(j-1)}}{r^{(j)}}$.  Letting $x=\ln\frac{r^{(j-1)}}{r^{(j)}}<0$, the desired inequality becomes $-x>1-e^x$, which is immediate.

For (\ref{eqn::usContinuum}), we would like to prove that $\alpha<\alpha^{(1)}$.  Note that $\alpha^{(1)}$ is the unique solution to
\begin{equation}\label{eqn::4941}
\alpha^{(1)}+\sum_{j=2}^m\Big[-\ln\big(1-(1-e^{-\alpha^{(1)}})(1-\frac{r^{(j-1)}}{r^{(j)}})\big)\Big]=1,
\end{equation}
while $\alpha$ is the unique solution to
\begin{equation}\label{eqn::4942}
\alpha+\sum_{j=2}^m(1-e^{-\alpha})\ln\frac{r^{(j)}}{r^{(j-1)}}=1.
\end{equation}

The LHS of (\ref{eqn::4941}), as a function of $\alpha^{(1)}$, is increasing over $(0,1)$; the same can be said about the LHS of (\ref{eqn::4942}) as a function of $\alpha$.  Therefore, it suffices to show that if $\alpha^{(1)}=\alpha=x$, then the LHS of (\ref{eqn::4941}) is strictly less than the LHS of (\ref{eqn::4942}), for all $x\in(0,1)$.

Let $F=1-e^{-x}$ and consider any $j>1$.  Let $s=\frac{r^{(j-1)}}{r^{(j)}}\in(0,1)$.  It suffices to show that $-\ln(1-F(1-s))<F\cdot\ln\frac{1}{s}$, which can be rearranged as $\frac{1-s^F}{1-s}>F$.  For the final inequality, note that $f(s)=s^F$ is a strictly concave function on $(0,1)$, since $F\in(0,1)$.  Therefore, $\frac{1-s^F}{1-s}>F$, because the LHS is the slope of the secant line through $(s,s^F)$ and $(1,1)$, while the RHS is the slope of the tangent line through $(1,1)$.
\Halmos\endproof

\section{Supplement to Section~\ref{sect::balance}}\label{appx::mr}

The first subsection contains the deferred proofs from Section~\ref{sect::balance}.  In the second subsection, we explain how to optimize the randomized procedure for generating a single value function.  In the third subsection, we put together the proof of Theorem~\ref{thm::mr}.

The following inequality will be useful throughout the paper.  For all $j=2,\ldots,m$, (\ref{eqn::solveAlpha}) says that $1-e^{-\alpha^{(j)}}\le1-\frac{r^{(j-1)}}{r^{(j)}}$, where we have used the fact that $1-e^{-\alpha^{(1)}}\le1$.  Therefore, for all $j=2,\ldots,m$, we can derive that
\begin{equation}\label{eqn::solveAlphaAgainUseful}
\frac{r^{(j-1)}}{r^{(j)}}\le e^{-\alpha^{(j)}}. 
\end{equation}

\subsection{Deferred Proofs}

\proof{Proof of Theorem~\ref{thm::randomizedBalance}.}
Define $N_{t,i}$ to be the algorithm's value for $N_i$ at the \textit{end} of time $t$ ($N_{0,i}$ is understood to be 0), for all $t\in[T]$ and $i\in[n]$.  For all $t\in[T]$, define $R_t=r^{(j^*_t)}_{i^*_t}$ and $Z_t=\tPhi_{i^*_t}(\tL^{(j^*_t)}_{i^*_t})-\tPhi_{i^*_t}(N_{i^*_t}/k_{i^*_t})$ if a sale was made during time $t$; define $R_t=Z_t=0$ otherwise.

Consider the solution to the dual LP (\ref{dual}) formed by setting $y_i=\bE[\tPhi_i(\frac{N_{T,i}}{k_i})]$ for all $i\in[n]$, and $z_t=\bE[Z_t]$ for all $t\in[T]$.  We claim that this solution is feasible.  The non-negativity constraint (\ref{dual::nonneg}) can be verified directly from the definitions.

Now, consider constraint (\ref{dual::feas}) for a fixed $t\in[T],i\in[n],j\in[m_i]$.  Given the initializations of $\tL^{(1)}_i,\ldots,\tL^{(m_i)}_i,\tPhi_i$ and the value of $N_{t-1,i}$, the algorithm will always make a decision during time $t$ which earns pseudorevenue whose conditional expectation is at least $p^{(j)}_{t,i}(\tPhi_i(\tL^{(j)}_i)-\tPhi_i(\frac{N_{t-1,i}}{k_i}))$, by definition (\ref{eqn::pseudorev_balance}).  Formally,
\begin{equation*}\label{eqn::condExpn}
\bE[Z_t|\tL^{(1)}_i,\ldots,\tL^{(m_i)}_i,\tPhi_i,N_{t-1,i}]\ge p^{(j)}_{t,i}(\tPhi_i(\tL^{(j)}_i)-\tPhi_i(\frac{N_{t-1,i}}{k_i})),
\end{equation*}
for all values of $\tL^{(1)}_i,\ldots,\tL^{(m_i)}_i,\tPhi_i,N_{t-1,i}$.  By the tower property of conditional expectation, $z_t=\bE[Z_t]\ge\bE[p^{(j)}_{t,i}(\tPhi_i(\tL^{(j)}_i)-\tPhi_i(\frac{N_{t-1,i}}{k_i}))]$.  Meanwhile, $y_i$ has been set to $\bE[\tPhi_i(\frac{N_{T,i}}{k_i})]$.  Since $N_{T,i}\ge N_{t-1,i}$ and $\tPhi_i$ is increasing, $y_i\ge\bE[\tPhi_i(\frac{N_{t-1,i}}{k_i})]$.  Therefore, the LHS of (\ref{dual::feas}), $p^{(j)}_{t,i}y_i+z_t$, is at least $\bE[p^{(j)}_{t,i}(\tPhi_i(\tL^{(j)}_i))]$.  By (\ref{eqn::feasibility}), this is at least $r^{(j)}_i$, completing the proof of feasibility.

Applying weak duality, we obtain
\begin{align}
\OPT(\cS,\cA) &\le\sum_{i=1}^nk_i\bE[\tPhi_i(\frac{N_{T,i}}{k_i})]+\sum_{t=1}^T\bE[Z_t] \nonumber \\
&=\sum_{i=1}^nk_i\bE\Big[\sum_{t=1}^T(\tPhi_i(\frac{N_{t,i}}{k_i})-\tPhi_i(\frac{N_{t-1,i}}{k_i}))\Big]+\sum_{t=1}^T\bE[Z_t] \nonumber \\
&=\sum_{t=1}^T\bE\Big[\sum_{i=1}^nk_i(\tPhi_i(\frac{N_{t,i}}{k_i})-\tPhi_i(\frac{N_{t-1,i}}{k_i}))+Z_t\Big]. \label{eqn::returnHere}
\end{align}
We now analyze the term inside the expectation,
\begin{equation}\label{eqn::6153}
\sum_{i=1}^nk_i(\tPhi_i(\frac{N_{t,i}}{k_i})-\tPhi_i(\frac{N_{t-1,i}}{k_i}))+Z_t,
\end{equation}
for every $t\in[T]$.  We would like to argue that it is at most $\frac{R_t}{c}$, on every sample path.

There are two cases.  If an item $i=i^*_t$ was sold at price $j=j^*_t$ during time $t$, then (\ref{eqn::6153}) equals
\begin{equation}\label{eqn::8945}
k_{i}(\tPhi_i(\frac{N_{t-1,i}+1}{k_i})-\tPhi_i(\frac{N_{t-1,i}}{k_i}))+\tPhi_i(\tL^{(j)}_{i})-\tPhi_i(\frac{N_{t-1,i}}{k_i}).
\end{equation}
Indeed, $N_{t,i}=N_{t-1,i}+1$, $N_{t,i}=N_{t-1,i}$ for all $i\neq i$, and $Z_t=\tPhi_i(\tL^{(j)}_{i})-\tPhi_i(\frac{N_{t-1,i}}{k_i})$ by definition.  Furthermore, since $Z_t$ is positive, $N_{t-1,i}$ must by less than $\tL^{(j)}_{i}k$.  Therefore, we can invoke (\ref{eqn::optimality}) to get that (\ref{eqn::8945}) is at most $r^{(j)}_{i}/c$, which is equal to $\frac{R_t}{c}$ by definition.  In the other case, if no item was sold during time $t$, then (\ref{eqn::8945}) is 0, while $R_t=0$ too, so (\ref{eqn::8945}) is still at most $\frac{R_t}{c}$.

Substituting back into (\ref{eqn::returnHere}), we conclude that $\OPT(\cS,\cA)\le\sum_{t=1}^T\bE[\frac{R_t}{c}]$, which is equal to $\frac{1}{c}\bE[\ALG(\cS,\cA)]$ by definition.  This completes the proof of Algorithm~\ref{alg::balance} having a competitive ratio at least $c$.
\Halmos\endproof

\proof{Proof of Proposition~\ref{prop::ppc}.}

For (\ref{eqn::ppc2}), note that $\bE[\tL^{(j)}]=\frac{\lfloor L^{(j)}k\rfloor+1}{k}(L^{(j)}k-\lfloor L^{(j)}k\rfloor)+\frac{\lfloor L^{(j)}k\rfloor}{k}(1-(L^{(j)}k-\lfloor L^{(j)}k\rfloor))=\frac{1}{k}(L^{(j)}k-\lfloor L^{(j)}k\rfloor)+\frac{\lfloor L^{(j)}k\rfloor}{k}=L^{(j)}$.

For (\ref{eqn::ppc3}), note that $|(\tL^{(j)}-\tL^{(j')})-(L^{(j)}-L^{(j')})|=|(\tL^{(j)}-L^{(j)})-(\tL^{(j')}-L^{(j')})|$.  We will prove that $(\tL^{(j)}-L^{(j)})-(\tL^{(j')}-L^{(j')})\le\frac{1}{k}$; the inequality that $(\tL^{(j)}-L^{(j)})-(\tL^{(j')}-L^{(j')})\ge-\frac{1}{k}$ follows by symmetry.  The maximum value of $k\tL^{(j)}$ is $\lfloor kL^{(j)}\rfloor+1$ while the minimum value of $k\tL^{(j')}$ is $\lfloor kL^{(j')}\rfloor$, hence the result is immediate unless $(\lfloor kL^{(j)}\rfloor+1)-kL^{(j)}+kL^{(j')}-\lfloor kL^{(j')}\rfloor>1$, i.e.\ $kL^{(j')}-\lfloor kL^{(j')}\rfloor>kL^{(j)}-\lfloor kL^{(j)}\rfloor$.  However, in this case, if $k\tL^{(j)}=\lfloor kL^{(j)}\rfloor+1$, then $W<kL^{(j)}-\lfloor kL^{(j)}\rfloor<kL^{(j')}-\lfloor kL^{(j')}\rfloor$ and hence $\tL^{(j')}$ is rounded up as well.  Similarly, if $\tL^{(j')}$ is rounded down, then $\tL^{(j)}$ must be rounded down as well.  If $\tL^{(j)}$ and $\tL^{(j')}$ are rounded in the same direction, then (iii) holds.
\Halmos\endproof

\proof{Proof of Theorem~\ref{thm::largeInv}.}
First we prove (\ref{eqn::feasibility}), the claim that $\bE[\tPhi(\tL^{(j)})]\ge r^{(j)}$, inductively.  Clearly $\bE[\Phi(\tL^{(0)})]\ge r^{(0)}=0$.  Now consider $j\in[m]$ and suppose we have established (\ref{eqn::feasibility}) for the $j-1$ case.  We can compare expression (\ref{eqn::tphi}) with $q=\tL^{(j)}$ and $q=\tL^{(j-1)}$ to obtain $\tPhi(\tL^{(j)})=\tPhi(\tL^{(j)})+(r^{(j)}-r^{(j-1)})\frac{\exp(\tL^{(j)}-\tL^{(j-1)})-1}{\exp(\alpha^{(j)})-1}$.  Therefore,
\begin{align*}
\bE[\tPhi(\tL^{(j)})] &\ge r^{(j-1)}+(r^{(j)}-r^{(j-1)})\frac{\bE[\exp(\tL^{(j)}-\tL^{(j-1)})]-1}{\exp(\alpha^{(j)})-1} \\
&\ge r^{(j-1)}+(r^{(j)}-r^{(j-1)})\frac{\exp(\bE[\tL^{(j)}-\tL^{(j-1)}])-1}{\exp(\alpha^{(j)})-1} \\
&=r^{(j-1)}+(r^{(j)}-r^{(j-1)})\frac{\exp(\alpha^{(j)})-1}{\exp(\alpha^{(j)})-1}
\end{align*}
where the first inequality uses the induction hypothesis, and the second inequality uses Jensen's inequality (the exponential function $\exp$ is convex).  The equality follows from (\ref{eqn::ppc2}) and the definition that $\alpha^{(j)}=L^{(j)}-L^{(j-1)}$, completing the induction.

Now we prove (\ref{eqn::optimality}) for an arbitrary $j\in[m]$ and $N\in\{0,\ldots,\tL^{(j)}k-1\}$.  Let $q=\frac{N}{k}$ and $\ell=\tell(q)$.  Note that $1\le\ell\le j$, and $\tL^{(\ell-1)}\le q<\tL^{(\ell)}$.  Substituting $q=\frac{N}{k}$ into the LHS of (\ref{eqn::optimality}), we get $k\big(\tPhi(q+\frac{1}{k})-\tPhi(q)\big)+\tPhi(\tL^{(j)})-\tPhi(q)$.  Adding and subtracting $\tPhi(\tL^{(\ell)})$ and rearranging, we get
\begin{equation}\label{eqn::4894}
k\big(\tPhi(q+\frac{1}{k})-\tPhi(q)\big)+\tPhi(\tL^{(\ell)})-\tPhi(q)+\tPhi(\tL^{(j)})-\tPhi(\tL^{(\ell)}).
\end{equation}
The following upper bound can be derived for expression (\ref{eqn::4894}):
\footnotesize
\begin{align}
&k\big(\tPhi(q+\frac{1}{k})-\tPhi(q)\big)+\tPhi(\tL^{(\ell)})-\tPhi(q)+\tPhi(\tL^{(j)})-\tPhi(\tL^{(\ell)}) \nonumber \\
= &(r^{(\ell)}-r^{(\ell-1)})\frac{e^{q+1/k-\tL^{(\ell-1)}}(k-(k+1)e^{-1/k})+e^{\tL^{(\ell)}-\tL^{(\ell-1)}}}{e^{\alpha^{(\ell)}}-1}+\sum_{\ell'=\ell+1}^j(r^{(\ell')}-r^{(\ell'-1)})\frac{e^{\tL^{(\ell')}-\tL^{(\ell'-1)}}-1}{e^{\alpha^{(\ell')}}-1} \nonumber \\
\le &(r^{(\ell)}-r^{(\ell-1)})\frac{e^{\tL^{(\ell)}-\tL^{(\ell-1)}}(k-(k+1)e^{-1/k})+e^{\tL^{(\ell)}-\tL^{(\ell-1)}}}{e^{\alpha^{(\ell)}}-1}+\sum_{\ell'=\ell+1}^j(r^{(\ell')}-r^{(\ell'-1)})\frac{e^{\tL^{(\ell')}-\tL^{(\ell'-1)}}-1}{e^{\alpha^{(\ell')}}-1} \nonumber \\
= &(r^{(\ell)}-r^{(\ell-1)})\frac{e^{\tL^{(\ell)}-\tL^{(\ell-1)}-\alpha^{(\ell)}}(1+k)(1-e^{-1/k})}{1-e^{-\alpha^{(\ell)}}}+\sum_{\ell'=\ell+1}^j(r^{(\ell')}-r^{(\ell'-1)})\frac{e^{\tL^{(\ell')}-\tL^{(\ell'-1)}-\alpha^{(\ell')}}-e^{-\alpha^{(\ell')}}}{1-e^{-\alpha^{(\ell')}}}. \label{eqn::rongLHS}
\end{align}
\normalsize
The inequality holds because $k-(1+k)e^{-1/k}>0$ for all $k\in\bN$, and $q$ is at most $\tL^{(\ell)}-1/k$.

It suffices to show that expression (\ref{eqn::rongLHS}) is bounded from above by
\begin{equation}\label{eqn::rongRHS}
r^{(j)}\frac{(1+k)(e^{1/k}-1)}{1-e^{-\alpha^{(1)}}}.
\end{equation}
To assist in this task, we would like to establish the following for all $\ell'=\ell+1,\ldots,j$ and $\ell''\in\{\ell,\ldots,\ell'-1\}$:
\footnotesize
\begin{align}
&(r^{(\ell'-1)}-r^{(\ell'-2)})\frac{e^{\tL^{(\ell'-1)}-\tL^{(\ell''-1)}-L^{(\ell'-1)}+L^{(\ell''-1)}}(1+k)(1-e^{-1/k})}{1-e^{-\alpha^{(\ell'-1)}}}+(r^{(\ell')}-r^{(\ell'-1)})\frac{e^{\tL^{(\ell')}-\tL^{(\ell'-1)}-\alpha^{(\ell')}}-e^{-\alpha^{(\ell')}}}{1-e^{-\alpha^{(\ell')}}} \nonumber \\
\le &(r^{(\ell')}-r^{(\ell'-1)})\frac{e^{\max\{\tL^{(\ell')}-\tL^{(\ell''-1)}-L^{(\ell')}+L^{(\ell''-1)},\tL^{(\ell')}-\tL^{(\ell'-1)}-L^{(\ell')}+L^{(\ell'-1)}\}}(1+k)(1-e^{-1/k})}{1-e^{-\alpha^{(\ell')}}}. \label{eqn::pieceInduction}
\end{align}
\normalsize
But $\frac{r^{(\ell'-1)}-r^{(\ell'-2)}}{1-e^{-\alpha^{(\ell'-1)}}}=\frac{r^{(\ell')}-r^{(\ell'-1)}}{1-e^{-\alpha^{(\ell')}}}\cdot\frac{r^{(\ell'-1)}}{r^{(\ell')}}$ due to the definition of $\alpha$ in (\ref{eqn::solveAlpha}), and $\frac{r^{(\ell'-1)}}{r^{(\ell')}}\le e^{-\alpha^{(\ell')}}$ due to (\ref{eqn::solveAlphaAgainUseful}).  Substituting back into inequality (\ref{eqn::pieceInduction}), it suffices to prove
\begin{align*}
&e^{\tL^{(\ell'-1)}-\tL^{(\ell''-1)}-L^{(\ell')}+L^{(\ell''-1)}}(1+k)(1-e^{-1/k})+e^{\tL^{(\ell')}-\tL^{(\ell'-1)}-\alpha^{(\ell')}}-e^{-\alpha^{(\ell')}} \nonumber \\
\le &e^{\max\{\tL^{(\ell')}-\tL^{(\ell''-1)}-L^{(\ell')}+L^{(\ell''-1)},\tL^{(\ell')}-\tL^{(\ell'-1)}-L^{(\ell')}+L^{(\ell'-1)}\}}(1+k)(1-e^{-1/k})
\end{align*}
where we have used Definition~\ref{defn::phi} to rewrite the first exponent.  Now, 
\begin{equation*}
e^{\tL^{(\ell'-1)}-\tL^{(\ell''-1)}-L^{(\ell')}+L^{(\ell''-1)}}(k-(1+k)e^{-1/k})\le e^{\tL^{(\ell')}-\tL^{(\ell''-1)}-L^{(\ell')}+L^{(\ell''-1)}}(k-(1+k)1-e^{-1/k}),
\end{equation*}
since $k-(1+k)e^{-1/k}>0$ and $\tL^{(\ell'-1)}\le\tL^{(\ell')}$.  Thus it remains to prove that
\footnotesize
\begin{equation}\label{eqn::considerCases}
e^{\tL^{(\ell'-1)}-\tL^{(\ell''-1)}-L^{(\ell')}+L^{(\ell''-1)}}+e^{\tL^{(\ell')}-\tL^{(\ell'-1)}-\alpha^{(\ell')}}-e^{-\alpha^{(\ell')}}\le e^{\max\{\tL^{(\ell')}-\tL^{(\ell''-1)}-L^{(\ell')}+L^{(\ell''-1)},\tL^{(\ell')}-\tL^{(\ell'-1)}-L^{(\ell')}+L^{(\ell'-1)}\}}.
\end{equation}
\normalsize
We consider two cases.  First suppose $\tL^{(\ell')}-\tL^{(\ell''-1)}-L^{(\ell')}+L^{(\ell''-1)}\le\tL^{(\ell')}-\tL^{(\ell'-1)}-L^{(\ell')}+L^{(\ell'-1)}$, i.e. $\tL^{(\ell'-1)}-\tL^{(\ell''-1)}\le L^{(\ell'-1)}-L^{(\ell''-1)}$.  Then the LHS of (\ref{eqn::considerCases}) equals $e^{-\alpha^{(\ell')}}+e^{\tL^{(\ell')}-\tL^{(\ell'-1)}-\alpha^{(\ell')}}-e^{-\alpha^{(\ell')}}=e^{\tL^{(\ell')}-\tL^{(\ell'-1)}-\alpha^{(\ell')}}$, which equals the RHS of (\ref{eqn::considerCases}) by the assumption that $\tL^{(\ell'-1)}-\tL^{(\ell''-1)}\le L^{(\ell'-1)}-L^{(\ell''-1)}$.  In the second case, suppose $\tL^{(\ell')}-\tL^{(\ell''-1)}-L^{(\ell')}+L^{(\ell''-1)}>\tL^{(\ell')}-\tL^{(\ell'-1)}-L^{(\ell')}+L^{(\ell'-1)}$, i.e. $\tL^{(\ell'-1)}-\tL^{(\ell''-1)}>L^{(\ell'-1)}-L^{(\ell''-1)}$.  Then inequality (\ref{eqn::considerCases}) can be rearranged as
\begin{equation*}
e^{-\alpha^{(\ell')}}(e^{\tL^{(\ell'-1)}-\tL^{(\ell''-1)}-L^{(\ell'-1)}+L^{(\ell''-1)}}-1)(e^{\tL^{(\ell')}-\tL^{(\ell'-1)}}-1)\ge0.
\end{equation*}
The first bracket is positive by the assumption that $\tL^{(\ell'-1)}-\tL^{(\ell''-1)}>L^{(\ell'-1)}-L^{(\ell''-1)}$ and the second bracket is non-negative since $\tL^{(\ell'-1)}\le\tL^{(\ell')}$.  This finishes the proof of (\ref{eqn::considerCases}), and hence (\ref{eqn::pieceInduction}).

Equipped with (\ref{eqn::pieceInduction}), we return the task of proving that expression (\ref{eqn::rongLHS}) is at most expression (\ref{eqn::rongRHS}).  If we inductively apply inequality (\ref{eqn::pieceInduction}) to expression (\ref{eqn::rongLHS}) for $\ell'=\ell+1,\ldots,j$ (when $\ell'=\ell+1$, $\ell''=\ell$; when $\ell'=\ell+2$, $\ell''=\ell$ if we arrived at case two during iteration $\ell+1$ and $\ell''=\ell+1$ otherwise,...), we conclude that expression (\ref{eqn::rongLHS}) is bounded from above by 
\begin{equation*}
(r^{(j)}-r^{(j-1)})\frac{e^{\tL^{(j)}-\tL^{(\ell''-1)}-L^{(j)}+L^{(\ell''-1)}}(1+k)(1-e^{-1/k})}{1-e^{-\alpha^{(j)}}}
\end{equation*}
for some $\ell''\in\{\ell,\ldots,j\}$.  The fact that $1-e^{-\alpha^{(1)}}=\frac{r^{(j)}}{r^{(j)}-r^{(j-1)}}(1-e^{-\alpha^{(j)}})$, due to (\ref{eqn::solveAlpha}), and the fact that $(\tL^{(j)}-\tL^{(\ell''-1)})-(L^{(j)}-L^{(\ell''-1)})\le1/k$, due to (\ref{eqn::ppc3}), complete the proof of expression (\ref{eqn::rongLHS}) being at most expression (\ref{eqn::rongRHS}), and thus the proof of Theorem~\ref{thm::largeInv} for general $m$.

Finally, when $m=1$, $\alpha^{(1)}=1$.  In the above proof, since $j$ and $\ell$ are always 1, (\ref{eqn::rongLHS}) can be replaced by $r^{(1)}\cdot\frac{(1)(1+k)(1-e^{-1/k})}{1-e^{-1}}$, where we have used the fact that $L^{(1)}=k$ always.  This is immediately at most $\frac{r^{(j)}}{c}$, for the improved value of $c=\frac{1-e^{-\alpha^{(1)}}}{(1+k)(1-e^{-1/k})}$, completing the proof of Theorem~\ref{thm::largeInv} in its entirety.
\Halmos\endproof

\subsection{Optimizing the Randomized Procedure}

We can explicitly formulate the optimization problem over randomized procedures for a single item with starting inventory $k$ and $m$ prices $r^{(1)},\ldots,r^{(m)}$.  Using the ``balls in bins'' counting argument, the number of configurations satisfying (\ref{eqn::random_segment_borders}) is $D:=\binom{k+m-1}{m-1}$.

We refer to these configurations in an arbitrary order using the index $d\in[D]$, where we let $\rho_d$ denote the probability of choosing configuration $d$, $f_d(\cdot)$ denote the value function for $d$, and $L^{(j)}_d$ denote the value of $\tL^{(j)}$ under configuration $d$ for all $j=0,\ldots,m$.  The optimization problem of satisfying (\ref{eqn::optimality})--(\ref{eqn::feasibility}) with a maximal value of $c$ can be formulated as follows:
\begin{subequations}\label{constr}
\begin{align}
\tCR:=\sup c & & \\
k(f_d(\frac{N+1}{k})-f_d(\frac{N}{k}))+f_d(L^{(j)}_d)-f_d(\frac{N}{k}) &\le\frac{r^{(j)}}{c} &d\in[D],j\in[m],0\le N\le kL^{(j)}_d-1 \label{constr::optimality} \\
f_d(1)\ge\ldots\ge f_d(\frac{1}{k})\ge f_d(0) &=0 &d\in[D] \label{constr::defnPen} \\
\sum_{d=1}^D\rho_df_d(L^{(j)}_d) &\ge r^{(j)} &j\in[m] \label{constr::feasibility} \\
\sum_{d=1}^D\rho_d &=1 & \label{constr::sumTo1} \\
f_d(0),f_d(\frac{1}{k}),\ldots,f_d(1)\in\bR;\rho_d &\ge0 & d\in[D] \label{constr::nonneg}
\end{align}
\end{subequations}
Constraint (\ref{constr::optimality}) corresponds to (\ref{eqn::optimality}), constraint (\ref{constr::feasibility}) corresponds to (\ref{eqn::feasibility}), while constraint (\ref{constr::defnPen}) enforces the definition of a value function in (\ref{eqn::random_value_fn}).  We let $\tCR$ denote the optimal objective value of (\ref{constr}).  Unfortunately, it is difficult to solve (\ref{constr}) exactly, since the number of configurations $D$ is exponential in the number of prices $m$, and constraint (\ref{constr::feasibility}) is non-linear.

Nonetheless, (\ref{constr}) is useful at determining the best competitive ratio which could be established \textit{using our analysis}.  We know that the randomized procedure from Definition~\ref{defn::tphi} (based on $\Phi$) is an optimal solution to (\ref{constr}) as $k\to\infty$, since it achieves the optimal competitive ratio possible.

We can also solve (\ref{constr}) exactly when $k=1$, in which case $D=m$, where we will let $d\in[D]$ denote the configuration with $\tL^{(0)}=\ldots=\tL^{(d-1)}=0$ and $\tL^{(d)}=\ldots=\tL^{(m)}=1$.  (\ref{constr::optimality}) reduces to $2f_d(1)\le\frac{r^{(j)}}{c}$, and needs to hold for $d\in[D]$, $j\ge d$ (for $j<d$, $kL^{(j)}_d-1=-1$).  However, clearly only the constraint with $j=d$ is binding.  As a result, (\ref{constr::optimality}) corresponds to $m$ constraints.  (\ref{constr::feasibility}) corresponds to $m$ constraints of the form $\sum_{d=1}^j\rho_df_d(1)\ge r^{(j)}$, for $j\in[m]$.

Not counting $f_d(0)$, which must be set to 0, there are $2m+1$ variables: $\{f_d(1),\rho_d:d\in[D]\}$ and $c$.  Consider the system of equations obtained in these $2m+1$ variables by setting (\ref{constr::optimality}), (\ref{constr::feasibility}), and (\ref{constr::sumTo1}) to equality.  It can be checked that the unique solution is
\begin{equation}\label{eqn::singleUnitSoln}
f_d(1)=\frac{r^{(d)}}{\sigma^{(1)}},\forall d\in[D];\rho_d=\sigma^{(d)},\forall d\in[D];c=\frac{\sigma^{(1)}}{2}
\end{equation}
with $\sigma^{(1)},\ldots,\sigma^{(m)}$ defined from $r^{(1)},\ldots,r^{(m)}$ according to (\ref{eqn::solveSigma}).  Furthermore, this solution is both feasible, satisfying the non-negativity constraints in (\ref{constr::defnPen}) and (\ref{constr::nonneg}), and optimal.  Therefore, the value of $\tCR$ is $\frac{\sigma^{(1)}}{2}$.

\subsection{Proof of Theorem~\ref{thm::mr}}

Now we put together the proof of Theorem~\ref{thm::mr}.  For all items $i\in[n]$, $\tCR_i$ is defined to be the optimal objective value of (\ref{constr}), with $k=k_i$, $m=m_i$, and $r^{(1)}=r^{(1)}_i,\ldots,r^{(m)}=r^{(m_i)}_i$.  Consider Algorithm~\ref{alg::balance}, where for all $i$, the randomized procedure used to initialize $\tPhi_i$ is an optimal solution to (\ref{constr}) achieving the objective value of $\tCR_i$.  For all $i$, (\ref{eqn::optimality})--(\ref{eqn::feasibility}) is satisfied as long as $c\le\tCR_i$.  Therefore, the maximum value of $c$ satisfying the conditions of Theorem~\ref{thm::randomizedBalance} is $\min_i\tCR_i$.  By Theorem~\ref{thm::randomizedBalance}, this algorithm achieves a competitive ratio of $\min_i\tCR_i$.

To establish bounds (i)--(iii) from Theorem~\ref{thm::mr}, for all $i$, we need to find a feasible randomized procedure with an objective value in (\ref{constr}) equal to the bound.  For bounds (i) and (iii), this is established directly by the randomized procedure from Definition~\ref{defn::tphi}, via the statement of Theorem~\ref{thm::largeInv}.  For bound (ii), we need to split the $k_i$ units of item $i$ into $k_i$ disparate items.  For each single-unit item, its value function in Algorithm~\ref{alg::balance} is initialized according to the randomized procedure described by (\ref{eqn::singleUnitSoln}).  This yields a value of $\frac{\sigma^{(1)}_i}{2}$, completing the proof of Theorem~\ref{thm::mr}.

\section{Deferred Proofs from Section~\ref{sect::ranking}}\label{appx::other}

\proof{Proof of Lemma~\ref{lem::djkOpt}.}
Since the algorithm was willing to sell item $i$ at price $j$, it must be the case that $W_i<L^{( j)}_i$.  Let $\ell$ denote $\ell_i(W_i)$, which is at most $j$.  We ignore measure-zero events and assume that $W_i\neq L^{(\ell-1)}$.  We can rearrange $Z_t$ as
\begin{align*}
&r^{(j)}_i-r^{(\ell)}_i+r^{(\ell)}_i-\Big(r^{(\ell-1)}_i+(r^{(\ell)}_i-r^{(\ell-1)}_i)\frac{\exp(W_i-L^{(\ell-1)}_i)-1}{\exp(\alpha^{(\ell)}_i)-1}\Big) \\
&=r^{(j)}_i-r^{(\ell)}_i+(r^{(\ell)}_i-r^{(\ell-1)}_i)\frac{\exp(\alpha^{(\ell)}_i)-\exp(W_i-L^{(\ell-1)}_i)}{\exp(\alpha^{(\ell)}_i)-1}.
\end{align*}
Adding $Y_i=\Phi'_{\cP_i}(W_i)=(r^{(\ell)}_i-r^{(\ell-1)}_i)\frac{\exp(W_i-L^{(\ell-1)}_i)}{\exp(\alpha^{(\ell)}_i)-1}$ to this expression, we get $r^{(j)}_i-r^{(\ell)}_i+\frac{r^{(\ell)}_i-r^{(\ell-1)}_i}{1-\exp(-\alpha^{(\ell)}_i)}$, which can be re-written as $r^{(j)}_i-r^{(\ell)}_i+\frac{r^{(\ell)}_i}{1-\exp(-\alpha^{(1)}_i)}$ due to (\ref{eqn::solveAlpha}).  The result follows immediately.
\Halmos\endproof

\proof{Proof of Lemma~\ref{lem::djkFeas}.}
It suffices to show that constraint (\ref{dual::feas}) holds for all $t\in[T]$ and $i\in[n]$.  Since $p^{(j)}_{t,i}\in\{0,1\}$ and the constraint clearly holds when $p^{(j)}_{t,i}=0$, it suffices to show that $\bE[Y_i+Z_t]\ge r^{(j_{t,i})}_i$, where $j_{t,i}\neq0$.  We will let $j=j_{t,i}$ for brevity.

Fix the realization of $W_{i'}$ for all $i'\neq i$, and consider the run of the algorithm on a modified setup with item $i$ removed.  Having fixed the values of $W_{i'}$, such a run is deterministic.  Let $\zcrit$ denote the pseudorevenue earned on this run during time $t$, possibly 0.  $\Phi_{\cP_i}$ maps $[0,L^{( j)}_i]$ to $[0,r^{(j)}_i]$ bijectively, so we can set $\wcrit$ to be the value in $[0,L^{(j)}_i]$ for which $\Phi_{\cP_i}(\wcrit)=\max\{r^{(j)}_i-\zcrit,0\}$.

We now consider the run of the algorithm on the full setup with item $i$, which is dependent on the realization of $W_i$.  The following two claims from \cite{DJK13} generalize to our multi-price setting.

\begin{enumerate}
\item Dominance: if $W_i\in[0,\wcrit)$, then in the run with item $i$, item $i$ gets matched.
\end{enumerate}
\underline{Proof}: Since $\wcrit>W_i$ and $W_i\ge0$, $\wcrit>0$.  Therefore, $\Phi_{\cP_i}(\wcrit)>0$.  Thus $\Phi_{\cP_i}(\wcrit)=r^{(j)}_i-\zcrit$ (as opposed to $\Phi_{\cP_i}(\wcrit)=0$), and moreover since $W_i<\wcrit$ and $\Phi_{\cP_i}$ is strictly increasing, $\Phi_{\cP_i}(W_i)<r^{(j)}_i-\zcrit$.  This implies $r^{(j)}_i-\Phi_{\cP_i}(W_i)>\max\{\zcrit,0\}$, since $\zcrit\ge0$.  Thus on the run with item $i$, either $i$ is already matched before time $t$, or it is matched to customer $t$.

\begin{enumerate}[resume]
\item Monotonicity: $Z_t\ge\zcrit$ (regardless of the realization of $W_i$).
\end{enumerate}
\underline{Proof}: fix the realization of $W_i$.  We compare two deterministic runs of the algorithm: one with item $i$, and one without.  We can inductively establish over $t=0,\ldots,T$ that at the end of time $t$, the set of unmatched items in the run with $i$ is a superset of that in the run without $i$.  Therefore, in the run with $i$, since the algorithm is maximizing pseudorevenue over a superset of items, its pseudorevenue $Z_t$ can be no less than $\zcrit$.

Now, conditioned on the realizations of $W_{i'}$ for $i'\neq i$, which determines the values of $\zcrit$ and $\wcrit$, we have $Z_t\ge\zcrit$ (by Monotonicity) and in turn $\zcrit\ge r^{(j)}_i-\Phi_{\cP_i}(\wcrit)$ (by the definition of $\wcrit$).  Meanwhile, as long as $i$ gets matched, $Y_i$ gets set to $\Phi'_{\cP_i}(W_i)$, so by Dominance, $\bE[Y_i|\{W_{i'}:i'\neq i\}]\ge\int_0^{\wcrit}\Phi'_{\cP_i}(w)dw=\Phi_{\cP_i}(\wcrit)-\Phi_{\cP_i}(0)=\Phi_{\cP_i}(\wcrit)$.  Therefore, $\bE[Y_i+Z_t|\{W_{i'}:i'\neq i\}]\ge r^{(j)}_i$.  The proof follows from the tower property of conditional expectation.
\Halmos\endproof

\section{Deferred Proofs from Section~\ref{sect::countereg}}

\proof{Proof of Proposition~\ref{prop::beta}.}
The unique solution to the system (\ref{eqn::propBeta}) is obtained inductively over $j=2,\ldots,m$ by setting $B_j=\frac{r^{(j-1)}e^{-\alpha^{(j-1)}}}{r^{(j)}e^{-\alpha^{(j)}}}B_{j-1}$.  By (\ref{eqn::solveAlphaAgainUseful}), $\frac{r^{(j-1)}}{r^{(j)}}\le e^{-\alpha^{(j)}}$, hence $B_j\le e^{-\alpha^{(j-1)}}B_{j-1}$.  But $\alpha^{(j-1)}>0$ by Proposition~\ref{prop::solveSys}, completing the proof that $B_j<B_{j-1}$ for $j=2,\ldots,m$.  The fact that $0<B_m$ is immediate.
\Halmos\endproof

\proof{Proof of Lemma~\ref{lem::ubDominance}.}
Consider the execution of an online algorithm with this randomized arrival sequence.  For all $i\in[n]$ and group of customers $t\in[n]$, let $Q_{t,i}$ denote the number of group-$t$ customers to which item $\pi_i$ is sold, which is a random variable with respect to the random permutation $\pi$ as well as any randomness in the algorithm.  Let $q_{t,i}=\bE[Q_{t,i}]$.

Clearly if $i<t$, then $Q_{t,i}=0$, because group-$t$ customers have no interest in item $\pi_i$.  Otherwise, for any $i,i'\ge t$, we argue that $q_{t,i}=q_{t,i'}$.  This is because while group $t$ is arriving, the online algorithm cannot distinguish between items $\pi_i$ and $\pi_{i'}$, hence any items it allocates are equally likely to be item $\pi_i$ and item $\pi_{i'}$.  Therefore, we let $q_t$ denote the value of $q_{t,i}$ for $i\ge t$.

Now, consider item $\pi_n$.  Since it only has $k$ units of inventory, we know that $\sum_{t=1}^nQ_{t,n}\le k$ on every sample path.  Using the linearity of expectation, we get that
\begin{equation}\label{eqn::msvvSumTo1}
\sum_{t=1}^nq_t\le k.
\end{equation}

Furthermore, for a $t\in[n]$, on every sample path, $\sum_{i=t}^nQ_{t,i}\le k$, since there are only $k$ customers in group $t$.  Therefore, $(n+1-t)q_t\le k$, or
\begin{equation}\label{eqn::harmonic}
q_t\le\frac{k}{n+1-t}.
\end{equation}

For this proof, let $M_j=\sum_{j'=1}^j\beta_{j'}$, for all $j=0,\ldots,m$.  For all $j\in[m]$, let $\lambda_j=\frac{1}{k}\sum_{t=M_{j-1}n+1}^{M_jn}q_t$.  Substituting into (\ref{eqn::msvvSumTo1}), we get the constraint that $\sum_{j=1}^m\lambda_j\le1$.  For any $j\in[m-1]$, summing inequality (\ref{eqn::harmonic}) for $t=M_{j-1}n+1,\ldots,M_jn$ yields $\lambda_j\le\ln\frac{B_j}{B_{j+1}}$, since $n\to\infty$, and $B_j=1-M_{j-1}$, $B_{j+1}=1-M_j$ by definition.  It is also clear from definition that $\lambda_j\ge0$ for all $j\in[m]$.

Finally, the total expected revenue is
\begin{equation}\label{eqn::totalLambda}
\sum_{j=1}^mr^{(j)}\sum_{t=M_{j-1}n+1}^{M_jn}q_t(n+1-t),
\end{equation}
since for each group $t$ there are $n+1-t$ items for each of which $q_t$ copies are sold in expectation.  Consider any $j\in[m]$.  Since $\sum_{t=M_{j-1}n+1}^{M_jn}q_t=\lambda_jk$ by definition, $\sum_{t=M_{j-1}n+1}^{M_jn}q_t(n+1-t)$ is maximized by setting $q_t$ to its upper bound in (\ref{eqn::harmonic}) for $t=M_{j-1}n+1,M_{j-1}n+2,\ldots$ until the capacity of $\lambda_jk$ is reached.  Since $n\to\infty$, we can simply compute the value of $t$ for which
\begin{equation}\label{eqn::7897}
\frac{k}{n-M_{j-1}n}+\ldots+\frac{k}{n-t}=\lambda_jk,
\end{equation}
with $t\in[M_{j-1}n,M_jn]$.  Letting $t=(M_{j-1}+y\beta_j)n$ with $y\in[0,1]$, and using the definition of $B_j$, (\ref{eqn::7897}) becomes $\ln\frac{B_j}{B_j-y\beta_j}=\lambda_j$, or $y\beta_j=B_j(1-e^{-\lambda_j})$.  Therefore,
\begin{align*}
\sum_{t=M_{j-1}n+1}^{M_jn}q_t(n+1-t) &\le\sum_{t=M_{j-1}n+1}^{(M_{j-1}+B_j(1-e^{-\lambda_j}))n}\frac{k}{n+1-t}\cdot(n+1-t) \\
&=B_j(1-e^{-\lambda_j})nk
\end{align*}
Substituting into (\ref{eqn::totalLambda}), we get that the expected revenue of the online algorithm is at most (\ref{eqn::ubALG}), where $\sum_{j=1}^m\lambda_j\le1$, $\lambda_j\le\ln\frac{B_j}{B_{j+1}}$ for $j\in[m-1]$, and $\lambda_j\ge0$ for $j\in[m]$, completing the proof.
\Halmos\endproof

\proof{Proof of Lemma~\ref{lem::valueFn}.}
We use backward induction over $j=m,\ldots,1$.  When $j=m$, (\ref{eqn::valueFn}) becomes $nkr^{(m)}B_m(1-\exp(-\tau))$, since $A_m=\alpha^{(m)}$ by definition.  Meanwhile, (\ref{eqn::backwardInduct}) is maximized by setting $\lambda_m=\tau$, resulting in the same expression and establishing the base case.

Now suppose $j<m$ and that we have already established the lemma in the $j+1$ case.  If we set $\lambda_j=\lambda$, for some $\lambda\in[0,\tau]$, then the maximum value of (\ref{eqn::backwardInduct}) subject to $\lambda_{j+1},\ldots,\lambda_m\ge0$ and $\lambda_{j+1}+\ldots+\lambda_m\le\tau-\lambda$ is, by the inductive hypothesis,
\begin{equation}\label{eqn::fnOfTau}
r^{(j)}B_j(1-\exp(-\lambda))nk+nk\sum_{\ell=j+1}^mr^{(\ell)}B_{\ell}\Big(1-\exp\big(-\alpha^{(\ell)}+\frac{A_{j+1}-(\tau-\lambda)}{m-(j+1)+1}\big)\Big).
\end{equation}
Consider this expression as a function of $\lambda$.  The derivative is
\begin{equation}\label{eqn::derivative}
r^{(j)}B_j\exp(-\lambda)nk+nk\sum_{\ell=j+1}^mr^{(\ell)}B_{\ell}\cdot\frac{-1}{m-j}\cdot\exp\big(-\alpha^{(\ell)}+\frac{A_{j+1}-(\tau-\lambda)}{m-j}\big)
\end{equation}
and the second derivative is clearly negative, so the function is concave.  Therefore, it is maximized by setting the derivative to 0.  By definition (\ref{eqn::propBeta}), $r^{(\ell)}B_{\ell}e^{-\alpha^{(\ell)}}$ is identical for all $\ell=j+1,\ldots,m$, and equal to $r^{(j)}B_je^{-\alpha^{(j)}}$.  Thus setting (\ref{eqn::derivative}) to 0 implies:
\begin{align*}
\exp(\alpha^{(j)}-\lambda) &=\frac{1}{m-j}\sum_{\ell=j+1}^m\exp\big(\frac{A_{j+1}-(\tau-\lambda)}{m-j}\big) \\
\alpha^{(j)}-\lambda &=\frac{A_{j+1}-(\tau-\lambda)}{m-j}.
\end{align*}
Rearranging and using the definition that $A_{j+1}=A_j-\alpha^{(j)}$, we get $\lambda=\alpha^{(j)}-\frac{A_j-\tau}{m-j+1}$.  Substituting this value of $\lambda$ into (\ref{eqn::fnOfTau}), the expression $\frac{A_{j+1}-(\tau-\lambda)}{m-(j+1)+1}$ is equal to $\frac{A_j-\tau}{m-j+1}$, hence (\ref{eqn::fnOfTau}) is equal to (\ref{eqn::valueFn}), completing the induction and the proof of the lemma.
\Halmos\endproof

\section{Deriving the Multi-price Value Function $\Phi_{\cP}$}\label{appx::deriving}

In this section we explain how we optimized the value function $\Phi_{\cP}$ for a given price set $\cP$, leading to the system of equations in (\ref{eqn::solveAlpha}), and the functional form in (\ref{eqn::phi}).  In Appendix~\ref{appx::continuum}, we use the same method to derive the optimal value function when the price of an item can take any value in the continuum $[\Rmin,\Rmax]$.

Consider constraints (\ref{eqn::optimality})--(\ref{eqn::feasibility}) in Theorem~\ref{thm::largeInv} for a single item with $k\to\infty$.  Let $w=\frac{N}{k}$, and we deterministically set $\tPhi$ to some $\Phi$.  The goal is to solve for the $\Phi$ which maximizes the value of $F$.

Observe that
\begin{equation*}
\lim_{k\to\infty}k(\Phi(\frac{N+1}{k})-\Phi(\frac{N}{k}))=\lim_{k\to\infty}\frac{\Phi(w+1/k)-\Phi(w)}{1/k},
\end{equation*}
which is equal to the derivative of $\Phi$ as $w$, by definition ($\Phi$ will end up not being differentiable on a discrete set of measure 0, which can be ignored).  Therefore, (\ref{eqn::optimality}) is equivalent to 
\begin{equation}\label{eqn::derivePhi}
\Phi'(w)-\Phi(w)\le r^{(j)}(\frac{1}{F}-1),
\end{equation}
and needs to hold for all $j\in[m],w\in[0,L^{(j)}]$.  For a fixed $w\in(L^{(j-1)},L^{(j)})$, (\ref{eqn::derivePhi}) needs to hold for all $j'=j,\ldots,m$, but is clearly binding when $j'=j$.  Therefore, it suffices to fix a $j\in[m]$ and consider (\ref{eqn::derivePhi}) when $w\in(L^{(j-1)},L^{(j)})$.

We should point out that this simplification via the ``binding'' argument is not possible for a finite $k$ and random $\tPhi$, because then (\ref{eqn::derivePhi}) becomes $\tPhi'(w)-\tPhi(w)\le\frac{r^{(j)}}{F}-\tPhi(L^{(j)})$, and the RHS in fact may not be increasing in $j$.
This is why we resort to first solving for $\Phi$ when $k\to\infty$ and then defining $\tPhi$ as a random perturbation of $\Phi$.

If we set (\ref{eqn::derivePhi}) to equality for some $j\in[m]$ and all $w\in(L^{(j-1)},L^{(j)})$, and solve the differential equation, we get that $\Phi(w)$ must be of the form $Ce^w-r^{(j)}(\frac{1}{F}-1)$ on $(L^{(j-1)},L^{(j)})$.  Setting $\Phi(L^{(j-1)})=r^{(j-1)}$ and $\Phi(L^{(j)})=r^{(j)}$, we obtain
\begin{align}
C &=\frac{r^{(j)}-r^{(j-1)}}{e^{L^{(j)}}-e^{L^{(j-1)}}}; \nonumber \\
F &=\frac{1}{1-\frac{r^{(j-1)}}{r^{(j)}}}\cdot(1-e^{-\alpha^{(j)}}). \label{eqn::CRonSegmentj}
\end{align}
The RHS of (\ref{eqn::CRonSegmentj}) is the largest value of $F$ which allows (\ref{eqn::derivePhi}) to hold on segment $j$.  It is dependent on $\alpha^{(j)}$, which is equal to $L^{(j)}-L^{(j-1)}$, the length of segment $j$.  For (\ref{eqn::derivePhi}) to hold on all segments $j\in[m]$, $F$ must be set to $\min_j\frac{1}{1-r^{(j-1)}/r^{(j)}}\cdot(1-e^{-\alpha^{(j)}})$.

Therefore, we would like to choose segment lengths $\alpha^{(1)},\ldots,\alpha^{(m)}$ summing to 1 to maximize the minimum $\frac{1}{1-r^{(j-1)}/r^{(j)}}\cdot(1-e^{-\alpha^{(j)}})$, which is accomplished by setting $\frac{1}{1-r^{(j-1)}/r^{(j)}}\cdot(1-e^{-\alpha^{(j)}})$ equal for all $j\in[m]$.  This yields the system of equations (\ref{eqn::solveAlpha}), and Proposition~\ref{prop::solveSys}.  The resulting value of $F$ is equal to $1-e^{-\alpha^{(1)}}$, since $r^{(0)}=0$.  The resulting value of $C$, when substituted into the equation for $\Phi(w)$ on each segment $(L^{(j-1)},L^{(j)})$, yields (\ref{eqn::phi}).

The derivation of $\Phi$ we just completed, starting from condition (\ref{eqn::derivePhi}), comes from our analysis of \balance.
We note that the exact same inequality (\ref{eqn::derivePhi}) can also be derived from our analysis of \ranking, which shows that the same value function should be used for both algorithms.

\subsection{Continuum of Feasible Prices}\label{appx::continuum}

Let the feasible price set for the item be $[\Rmin,\Rmax]$, where $0<\Rmin<\Rmax$.  Using the same ``binding'' argument, it suffices to maximize the value of $F$ for which the following can hold:
\begin{align}
\Phi'(w)-\Phi(w) &\le\Rmin(\frac{1}{F}-1), && w\in(0,\alpha); \label{eqn::contFirstPt} \\
\Phi'(w)-\frac{\Phi(w)}{F} &\le0, && w\in(\alpha,1). \label{eqn::contSecondPt}
\end{align}
$\Phi$ must also satisfy $\Phi(0)=0,\Phi(\alpha)=\Rmin,\Phi(1)=\Rmax$, while $\alpha\in(0,1)$ is an arbitrary ``booking limit'' for the lowest price of $\Rmin$.

We know from before that under the optimal solution to (\ref{eqn::contFirstPt}), the value of $F$ can be at most $1-e^{-\alpha}$.  Solving the differential equation where (\ref{eqn::contSecondPt}) is set to equality, $\Phi(w)$ must take the form $Ce^{w/F}$ on $(\alpha,1)$.  Substituting $\Phi(\alpha)=\Rmin$ and $\Phi(1)=\Rmax$ yields
\begin{align*}
C &=(\Rmin)^{\frac{1}{1-\alpha}}(\Rmax)^{-\frac{\alpha}{1-\alpha}}; \\
F &=\frac{1-\alpha}{\ln\frac{\Rmax}{\Rmin}}.
\end{align*}
Therefore, the value of $F$ is also bounded from above by $\frac{1-\alpha}{\ln(\Rmax/\Rmin)}$.  $F$ is maximized by setting $\frac{1-\alpha}{\ln(\Rmax/\Rmin)}$ equal to the other upper bound of $1-e^{-\alpha}$; the value at which equality is achieved is then the competitive ratio.

Letting $R=\ln(\Rmax/\Rmin)$, the solution to $\frac{1-\alpha}{R}=1-e^{-\alpha}$ can be written as $W(Re^{R-1})-R+1$, where $W$ is the Lambert-W function, the inverse function to $f(x)=xe^x$ for $x\in\bR_{\ge0}$.  Indeed, when $\alpha=W(Re^{R-1})$, the following can be derived:
\begin{align*}
\frac{1-\alpha}{R} &=1-e^{-\alpha} \\
Re^{-\alpha} &=\alpha+R-1 \\
Re^{R-1} &=(\alpha+R-1)e^{\alpha+R-1} \\
W(Re^{R-1}) &=\alpha+R-1
\end{align*}

Substituting $\alpha=W(\ln(\Rmax/\Rmin)e^{\ln(\Rmax/\Rmin)-1})-\ln(\Rmax/\Rmin)+1$ into the formula for $C$, and using the fact that $\Phi(w)=Ce^{w/F}$, we get
\begin{align*}
\Phi(w)=(\Rmin)^{\frac{1-w}{1-\alpha}}(\Rmax)^{\frac{w-\alpha}{1-\alpha}}, && w\in[\alpha,1].
\end{align*}
Meanwhile, the earlier derivation implies that
\begin{align*}
\Phi(w)=\Rmin\cdot\frac{e^w-1}{e^\alpha-1}, && w\in[0,\alpha].
\end{align*}
It can be checked that indeed $\Phi(0)=0$, $\Phi(\alpha)=\Rmin$ ($\Phi$ is continuous at $w=\alpha$), and $\Phi(1)=\Rmax$.  Furthermore, unlike the case of discrete prices, it can be checked that $\Phi$ is also differentiable at $w=\alpha$ (on $[\alpha,1]$, use the form that $\Phi(w)=Ce^{w/F}$, hence $\Phi'(\alpha)=\frac{\Phi(\alpha)}{F}$).

\section{Supplement to Numerical Experiments}\label{appx::numerical}

We provide additional details about our choice estimation.  We define 8 customer types, one for each combination of the 3 following binary features.
\begin{enumerate}
\item Group: whether the customer indicated a party size greater than 1.
\item CRO: whether the customer booked using the Central Reservation Office, as opposed to the hotel's website or a Global Distribution System (for details on these terms, see \cite{BFG09}).
\item VIP: whether the customer had any kind of VIP status.
\end{enumerate}
We did not use features such as: whether the booking date is a weekend, whether the check-in date is a weekend, the length of stay, or the number of days in advance booked.  Such features did not result in a more predictive model.

We estimate the mean MNL utilities for each of the 8 products separately for each customer type.  The results are displayed in Table~\ref{tbl::estimation}.  The total share of each customer type (out of all the transactions) is also displayed.  We should point out that it is possible for a customer to choose the higher fare for a room, even if the lower fare was also offered.  This is because the higher fares are often packaged with additional offers, such as airline services, city attractions, in-room services, etc.

\begin{table}
	\scriptsize
	\TABLE
	{MNL choice models for the 8 customer types.  The suffix ``L'' on a room type means lower fare, while the suffix ``H'' on a room type means higher fare.\label{tbl::estimation}}
	{\begin{tabular}{|ccc|c|ccccccccc|}
	\cline{1-3}\cline{5-13}
	\multicolumn{3}{|c|}{Customer Type} &  & \multicolumn{9}{c|}{MNL Mean Utilities} \\
		\hline
		 Group? & CRO? & VIP? & Share & KingL & QueenL & SuiteL & 2DoubleL & KingH & QueenH & SuiteH & 2DoubleH & NoBuy \\
		\hline
		 &  &  & 0.16 & -0.36 & -1.22 & -2.56 & -1.04 & 0 & -0.23 & -2.25 & -1.8 & 0 \\
		 &  & \checkmark & 0.03 & -0.82 & -1.98 & -2.16 & -2.09 & 0 & -1.02 & -1.45 & -1.82 & 0 \\
		 & \checkmark &  & 0.28 & -1.67 & $-\infty$ & -3.78 & -2.71 & 0 & -1.33 & -1.8 & -1.58 & 0 \\
		 & \checkmark & \checkmark & 0.09 & -2.13 & $-\infty$ & -3.38 & -3.76 & 0 & -2.12 & -1 & -1.59 & 0 \\
		\checkmark &  &  & 0.19 & -0.54 & -0.97 & -2.26 & 0 & -0.91 & -1.47 & -2.78 & -1.41 & 0 \\
		\checkmark &  & \checkmark & 0.04 & -0.09 & -0.82 & -0.95 & -0.14 & 0 & -1.35 & -1.07 & -0.51 & 0 \\
		\checkmark & \checkmark &  & 0.18 & -0.93 & $-\infty$ & -2.56 & -0.76 & 0 & -1.66 & -1.41 & -0.27 & 0 \\
		\checkmark & \checkmark & \checkmark & 0.03 & -1.39 & $-\infty$ & -2.16 & -1.8 & 0 & -2.45 & -0.61 & -0.28 & 0 \\
		\hline
	\end{tabular}}
	{}
\end{table}

We have shifted the mean utilities so that for each customer type, the weights of both the no-purchase option, and the most-preferred purchase option, is equal to 0.  (We synthetically set the weight of the no-purchase option because it is not possible to estimate from the data.)  The large weights on the no-purchase options ensure that the revenue-maximizing assortments tend to include both the low and high fares.

In the setting with greater fare differentiation (Subsection~\ref{subsect::fareDiff}), the high prices of the King, Queen, Suite, and Two-double rooms are adjusted to \$614, \$608, \$768, \$612, respectively (twice the lower fares).  The mean utility of the no-purchase option is increased by 2 for every customer type, to ensure that the revenue-maximizing assortments still include both the low and high fares.

\subsection{Details on the Forecasting Bid-price Algorithms}\label{appx::bidPrice}

To forecast the remaining number of customers, we assume that we know the average number of customers interested in each occupancy date (1340), as well as the overall trend for how far in advance customers book, which is plotted in Figure~\ref{fig::forecastCurve}.  As an example of how to use these numbers, consider the occupancy date March 31st.  At the start, we forecast there to be 1340 arrivals.  However, suppose by March 6th, 500 customers have arrived.  Since we know from Figure~\ref{fig::forecastCurve} that roughly 50\% of the total population interested in March 31st will have already booked by March 6th (25 days in advance), we expect there to only be 500 customers remaining.
\begin{figure}[t]
\begin{center}
\includegraphics[width=0.6\textwidth]{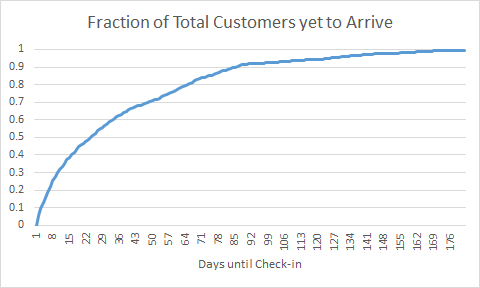}
\caption{Distribution of arrivals over the days before check-in, formed by aggregating all transactions.}\label{fig::forecastCurve}
\end{center}
\end{figure}

To forecast the breakdown of remaining customers by type, we assume that we know the aggregate distribution of customer type over all occupancy dates.  For example, from Appendix~\ref{appx::numerical}, we know that 28\% of all customers are of Type~3.  Then we would estimate $28\%\times500=140$ of the 500 remaining customers to be of Type~3.  Alternatively, one can try to learn the specific distribution of customers interested in March 31st.  Suppose that only 100, or 20\%, of the 500 bookings made before March 6th came from customers of Type~3.  Then we would instead estimate $20\%\times500=100$ of the 500 remaining customers to be of Type~3.

To use the forecasted information, algorithms incorporate it into the LP (\ref{primalAssortment}), and set the \textit{bid price} of each item $i$ equal to the shadow price of constraint $i$ in (\ref{primalAssortment::inv}).  These algorithms then offer each customer $t$ the assortment $S$ (from the available items) maximizing $\sum_{(i,j)\in S}p^{(j)}_{t,i}(S)\big(r^{(j)}_i-\lambda_i\big)$.

We clarify the exact way in which the forecasted information is incorporated into the LP.  Let there be $A$ customer types, indexed by $a=1,\ldots,A$.  We use $p^{(j)}_{a,i}(S)$ to denote the probability of a customer of type $a$ choosing product $(i,j)$ from assortment $S$.  Suppose that when we want to re-solve the LP (\ref{primalAssortment}), the forecasted number of remaining customers of type $a$ is $N_a$, for all $a\in[A]$, and the remaining inventory of item $i$ is $K_i$, for all $i\in[n]$.  We can formulate the following LP, which is a modification of (\ref{primalAssortment}):
\begin{align*}
\max\sum_{a=1}^A\sum_Sx_a(S)\sum_{(i,j)\in S}r^{(j)}_ip^{(j)}_{a,i}(S) & & \\
\sum_{a=1}^A\sum_Sx_a(S)\sum_{j:(i,j)\in S}p^{(j)}_{a,i}(S) &\le k_i &i\in[n] \\
\sum_Sx_a(S) &=N_a &a\in[A] \\
x_a(S) &\ge0 &a\in[A],S\subseteq\{(i,j):i\in[n],j\in[m_i]\}
\end{align*}
We have set $T=\sum_{a=1}^AN_a$ and $|\{t:\text{type of customer $t$ is $a$}\}|=N_a$; note that the ordering of remaining customers is inconsequential for the LP.

Although this LP has an exponential number of variables, we can easily solve it using column generation (e.g., see \cite{LvR08}).  Fix an optimal primal solution $(x^*_a(S):a\in[A],S\subseteq\{(i,j):i\in[n],j\in[m_i]\})$ and an optimal dual solution $(y^*_i:i\in[n]),(z^*_a:a\in[A])$.  The bid-price algorithm sets the bid price of each item $i$ equal to $y^*_i$.

We should point out that for every bid-price algorithm based on dual variables, there is a corresponding \textit{random assignment} algorithm based on primal variables.  Such an algorithm would, for each customer type $a$, offer each assortment $S$ with probability proportional to $x^*_a(S)$.  We have confirmed that these algorithms perform similarly in the simulations.  We compare with the bid-price algorithms instead of the random assignment algorithms because they follow a form more similar to our \balance algorithm.

\end{APPENDICES}







\end{document}